\documentclass{llncs}

\usepackage{url}
\usepackage{amsmath}
\usepackage{amssymb}

\usepackage{ifpdf}
 \ifpdf
\usepackage[pdftex]{graphicx}
\newcommand{\pix}{eps}% for pdf B&W, for new MikTex PDFlatex and ARXIV but only BLACK and WHITE =BW, avoid bb also
 \else
\usepackage[dvips]{graphicx}
\newcommand{\pix}{eps}% for pdf B&W, for new MikTex PDFlatex and ARXIV but only BLACK and WHITE =BW, avoid bb also
 \fi
\usepackage{pbox}%%for making line breaks inside tables!!!

\def\url#1{{\tt{#1}}}

\def\Z{\mbox{\rm{Z}\hspace{-.28em}\rm{Z}}}

\renewcommand{\thefigure}{\arabic{figure}}

\begin{document}
%they don't behave as \newtheorem{p}{P}[subsubsection], as they increase the counter [subsubsection] itself !!!
\spnewtheorem{theor}[subsection]{Theorem}{\bfseries}{\rmfamily}
\spnewtheorem{coro}[subsubsection]{Corollary}{\bfseries}{\rmfamily}
\spnewtheorem{defi}[subsection]{Definition}{\bfseries}{\rmfamily}
%\spnewtheorem{fact}[subsection]{Fact}{\bfseries}{\rmfamily}
\spnewtheorem{lem}[subsection]{Lemma}{\bfseries}{\rmfamily}

%\def\mytitl{
%On Lack of Unique Factorization
%%Events
%and a Construction of
%a Polynomial Invariant Attack of Degree 7 for a Block Cipher}
%Failure
\def\mytitl{Lack of Unique Factorization as a Tool in Block Cipher Cryptanalysis}
\title{\mytitl}
%\title{\mytitl \mytitm}
%
%\titlerunning{\mytit}
%\toctitle{\mytit}

\author{
Nicolas T. Courtois, Aidan Patrick
}
\authorrunning{N. T. Courtois}
%\tocauthor{\protect{}}

\institute{
University College London, Gower Street, London, UK
}

\maketitle
%\centerline{\Large \bf \mytitl}
%\vskip3pt
%\centerline{\Large \bf \mytitm}
%\addcontentsline{toc}{chapter}{Abstract}

\vskip-10pt
\vskip-10pt
\begin{abstract}
Linear (or differential) cryptanalysis may seem dull topics for a mathematician:
they are about super simple invariants characterised by say a word on $n=64$ bits with very few bits at 1,
the space of possible attacks is small, and basic principles are trivial.
In contract mathematics offers an infinitely rich world of possibilities. % which could make cryptographers pale.
If so, why is that cryptographers have ever found so few attacks on block ciphers?
In this paper we argue that black-box methods used so far to find attacks in symmetric cryptography are inadequate
and we work with a more recent white-box algebraic methodology.
Invariant attacks can be constructed explicitly through
the study of roots of the so-called Fundamental Equation (FE).
We also argue that certain properties of the ring of Boolean polynomials such as lack of unique factorization
allow for a certain type of product construction attacks to flourish.
As a proof of concept we show how to construct a complex and non-trivial attack
where a polynomial of degree 7 is an invariant
for any number of rounds for a complex block cipher. % T-310.
\vskip-3pt
\vskip-3pt
\end{abstract}

\vskip-1pt
\vskip-1pt
%\vskip-5pt
\vskip-5pt
\noindent
{\bf Key Words:~}
%%% full version maybe %%%
%later
%Cold War,
unique factorization,
cyclotomic integers,
algebraic number theory,
multivariate polynomials,
polynomial invariants,
irreducible polynomials,
prime numbers,
%triangular systems of equations,
%I/O sums,
block ciphers,
Boolean functions, non-linearity,
annihilator space,
ANF, polynomial rings
%Algebraic Normal Form,
%unbalanced
Feistel ciphers,
weak keys,
backdoors,
%compressing Feistel ciphers,
%history of cryptography,
T-310,
%Linear Cryptanalysis,
Generalized Linear Cryptanalysis,
%Partitioning Cryptanalysis,
%symmetric polynomials,
algebraic cryptanalysis.

\vskip-6pt
\vskip-6pt
\section{Introduction - Maths vs. Crypto Question}
\vskip-6pt

A major topic in number theory is to try to solve the usual problems by creating new abstract algebraic
extensions (or new types of imaginary objects), which extend the usual possibilities of doing arithmetic.
Solving diophantine equations and proving results about their solvability, is one of the
biggest research topics of all times, and have benefited greatly from various algebraic innovations
such as cyclotomic integers or various extension fields.
In cryptography the problem is similar:
breaking ciphers may be seen as solving a system of equations,
and finding new non-trivial cryptographically significant properties of ciphers can,
and should be formalised in this way.
An interesting area is block cipher cryptanalysis.
For decades the only attacks ever found were simple iterative attacks such as Linear Cryptanalysis
where the set of possible attacks is very small.
The space of possible attacks is tiny and it is hard to overlook the attacks if they exist.
Cryptanalysis with multivariate polynomials was studied extensively too \cite{desalg} but with very poor results.
We need to ask the following question:
why is it that cryptographers have ever found so few attacks.
Have mathematics have failed in its core mission of providing tools for understanding the world in which we live?
We hypothesise that the fact why we have never found many powerful attacks on block ciphers has three core reasons:

\vskip-5pt
\vskip-5pt
\begin{enumerate}
\item
One reason is combinatorial complexity: the number of all possible
non-linear polynomial invariant attacks grows as $2^{2^n}$.
Many authors stress that systematic exploration is not feasible \cite{BeiCantResNL}.
For this reason, there are extremely few positive results on this topic \cite{TodoNL18,BackdTut}
and any method to approximate the solution is valuable.
One new working example as in this paper will typically allow the researchers to find more similar examples.
\item
A second answer lies in the limitations of block-box work methodology (e.g. in Linear Cryptanalysis):
from basic properties on some black boxes we derive complex ones on larger boxes.
In this paper we hypothesise that cryptographers have been missing a lot by using this methodology.
We need a {\bf white-box} or algebraic methodology for the study of block ciphers and
once we have one, we are going to discover %a plethora of
new attacks never seen before.
\item
A third answer is about phase transitions and entry barriers.
A very substantial effort to find some attacks with lower degree polynomials failed to produce
anything more than just slightly better than best Matsui's attack, cf. \cite{BLC}.
A radical improvement can be achieved by switching to higher degree polynomial attacks
as shown in a recent paper \cite{BackdAnn}.
At a higher degree we suddenly discover that invariants true with probability 1 may exist.
\end{enumerate}
\vskip-5pt

In this paper we construct just one new attack on one block cipher setup.
%later after Kummer
Our invariant is a product of polynomials as in \cite{BackdAnn}.
However the steps needed to show that the attack works are more complex than ever before
and the attack does not resemble any previous attack.
Before we study further details, we are now going to argue that there exist an interesting property of polynomial arithmetic
which has been a source of serious trouble for mathematicians but
it is extremely helpful\footnote{
Our later attack is about equality of two products.}
%which explicitly
%We claim that it helps both in terms of probability
%and in how actually attacks can be discovered and studied.}
for finding new attacks on block ciphers.
%A proof of concept invariant attack will follow.

\vskip-6pt
\vskip-6pt
\subsection{Invention of Polynomial Rings vs. Unique Factorization}
\label{LameKummer}
\vskip-5pt

As already explained mathematics and number theory have developed by study and creation
of increasingly complex algebraic structures which were created as tools to solve important problems.
Rings are also useful in cryptography:
study of specific polynomial rings is at the heart of so called algebraic cryptanalysis.
Mathematics have over the centuries seen the role of abstract algebra increase.
The same is likely to happen if we want to make some progress in cryptanalysis.

%In 1847; Gabriel Lamé has submitted to the 19th century, the French academy of Science had offered a series of prices for proving last Fermat's theorem.
%withdrawn 1856
In 1847 Gabriel Lam\'{e} has announced that he has proven the Fermat's last theorem in the general case.
Actually two mathematicians: Lam\'{e} and Cauchy have submitted
so called ``secret packets'' recorded in the proceedings
of the Paris Academy of Science \cite{CyclotomicLackUFT}
which contained similar ideas.
The proof attempts were then reviewed and studied.
The work used the so called cyclotomic integers in $\Z[\alpha]$
%which can be viewed as polynomials with integer coefficient
%with one special variable being in
with a root of unity $\alpha$ with $\alpha^{p}=1$ and with $p=23$.
However Kummer has proven few years earlier that for $p=23$ %-th roots of unity the
ring $\Z[\alpha]$ does NOT have a unique factorization.
%47 has 2 factorizations! see http://fermatslasttheorem.blogspot.com/2006/07/cyclotomic-integers-unique_25.html
Thus a major attempt to prove the Fermat last theorem has failed,
and we had to wait for another 138 years for this question to be solved.

%\vskip-6pt
%\vskip-6pt
\subsection{Lack of Unique Factorization}
\vskip-4pt

In this paper we show that {\bf lack of unique factorization}
for multivariate Boolean polynomials has important consequences in block cipher cryptanalysis.
We claim that {\bf it helps the attacker} in a very strong way.
This is extremely clear here because we explore the so called ``product construction''
for polynomial invariant attacks on block ciphers proposed in \cite{BackdAnn}.
Without unique factorization there are substantially fewer (and only trivial) ways to make two polynomials
which are both products of linear factors equal to each other, which will be essentially permutations in the order of factors.
In this paper however we show that there is another way to make them equal. %method where factors are not permuted(!).
In general we conjecture that potentially every non-trivial polynomial invariant attack will contain something which are
%we could call
``lack of unique factorization events'' or annihilation events, absorbtion events etc, see
Thm. \ref{ProductEightThm26XGenThm7ver11} and page \pageref{LackUniqueFactUsed26511}
in Section \ref{Constr26XByHandStage1ProductAllDeg7ver11}.
An illuminating example of how this works can be found in page \pageref{LackUniqueFactUsed26511}
where a polynomial of degree 5 can be expressed in 2 different ways as a multiple
of two very different degree 3 polynomials where sets of variables are almost entirely disjoint (!),
cf. page \pageref{LackUniqueFactUsed26511}.

Recent papers show that with a lot of pain, one can construct some small degree invariant attacks on a block cipher with
invariants being of low degree \cite{BackdTut} and such attacks only work for extremely few degenerate and super weak Boolean functions.
Then the situation improves dramatically in \cite{BackdAnn} at degree 8. We get a ``phase transition'' from hard to easy,
or from a case where extremely few Boolean functions work, to where a very large proportion of Boolean functions will work with the attack.
The product attack constructed in \cite{BackdAnn} is a degree 8 invariant and product of 8 linear factors.
%where two polynomials are annihilated independently of each other and therefore a very complex polynomial can be annihilated.
In this paper we argue that a strong phase transition happens somewhere near degree 7 or 8, at least here
in this specific cipher setup.
However finding one strong attack does not solve the general problem HOW to find further attacks on block ciphers
(or attacks on stronger block ciphers). Here mathematics does not help a lot:
mathematicians have rarely studied situations which are abundant in cryptography where polynomials have
a large number of variables over a very small finite field.
A huge problem for a cryptanalyst here is combinatorial explosion or vast complexity:
the number of all possible polynomial invariant attacks on a cipher with $n$-bit blocks grows as $2^{2^n}$.
In this paper we argue that this problem can be approached by explicit white-box constructions of polynomial
invariant attacks where the attacker is able to make some immensely
complex polynomial reduce to zero: i.e. equal to zero for any of $2^{n}$ inputs.
It is also about tricks where the number of variables involved at some place can be reduced by some sort of miracle
and very complex polynomials get simply cancelled. %or annihilated.
Only recently such attacks have been show to exist
%Such attack were believed rather infeasible in cryptanalysis until recently,
%now we have numerous examples in a variety of shapes,
cf. \cite{BackdTut,TodoNL18}.

In this paper we construct a novel degree 7 invariant attack.
We prove step by step that our attack works.
%%already above
%Our proof is based on a key event where one single polynomial of degree 5 can be expressed in 2 different ways as a multiple
%of two completely different degree 3 polynomial where the sets of variables are almost entirely disjoint,
%cf. page \pageref{LackUniqueFactUsed26511}.
This attack is irregular and has a very different structure than any previous attack ever seen in block cipher cryptanalysis.
It does seem natural in any way that an attack with 7 well-chosen polynomials is at all possible for a highly regular Feistel cipher with 4 branches
and 4 internal stages inside the round function.
%We find these mathematical facts surprising.
We believe that this paper is a first step towards building a theory of algebraic polynomial invariant attacks on block ciphers.
We claim that possibly the only way to find some cryptographic attacks is to construct some attacks explicitly,
using specific algebraic identities,
even if this is done one weaker cipher or toy ciphers.
The dominant black-box methodology of finding new attacks by composition
should be considered as bankrupt, and for many decades cryptography researchers
have never found any complex polynomial attacks, only very simple ones.
This paper also suggest that the amount of mathematics and algebra in future works on block cipher cryptanalysis
will need to increase very substantially.

Our methodology for discovering the attacks is based
on the study of roots of the so-called Fundamental Equation (FE)
which is a simple
I/O sum of two polynomials (an Input polynomial and an Output polynomial)
and is formally defined much later in Section \ref{FEdef}.
Several methods for constructing polynomial invariant attacks and therefore also solving the $FE$
have been proposed recently.
One is by study of certain type of ``imperfect'' cycles on monomials cf. \cite{MariosMSc,BackdTut}.
Another is by RowEchelon elimination in polynomial spaces \cite{ConstrRowEchKT1,BackdTut}.
Many attacks have been constructed by paper and pencil methods \cite{MariosMSc,BackdTut,ConstrRowEchKT1,BackdAnn,InvHopWCC}.
Attacks can be transposed from one place to another \cite{BackdTut,ConstrRowEchKT1} or
applied to a modified cipher \cite{InvHopWCC}.
Finally we have a product construction in \cite{BackdAnn}.
The attack is constructed by multiplying polynomials from simpler attacks
where the equations have no roots (the do not correspond to any actual attack).
%Therefore
Yet eventually and suddenly we find ourselves at another, brighter side:
we construct an attack where the equation has roots \cite{InvHopWCC}
or even a large number of high quality roots \cite{BackdAnn}.

\vskip-6pt
\vskip-6pt
\section{Block Ciphers and Round Invariant Attacks}
\vskip-6pt

Block ciphers are in widespread use since the 1970s.
Their iterated structure is prone to numerous round invariant attacks
for example in Linear Cryptanalysis (LC).
The next step is to look at non-linear polynomial invariants with
%This is usually known as
Generalised Linear Cryptanalysis (GLC)
first proposed by Harpes, Kramer, and Massey
cf. \cite{GenLinear1} (Eurocrypt'95).
%%a more general approach introduced by Harpes and Massey in 1990s is
%%and Partitioning Cryptanalysis (PC).%%%cf. FSE'97.
%A major open problem in cryptanalysis is discovery
%of invariant properties of complex type.
%%%%below cf. for example polynomial invariants in two recent papers \cite{TodoNL18,BackdTut}.
%Researchers have until 2018 found extremely few such attacks
%with some impossibility results \cite{BeiCantResNL,FiliolNotVuln}.
%Eventually recent papers \cite{TodoNL18,BackdTut,MariosMSc,ConstrRowEchKT1} show how to construct polynomial invariant attacks for block ciphers,
%however in almost all such results the Boolean function is extremely weak %\cite{BackdTut,MariosMSc,ConstrRowEchKT1}
%and invariants are simple and of low degree
%%\cite{TodoNL18,BackdTut,MariosMSc,ConstrRowEchKT1}.
%Can we do better?
%What kind of better outcomes can we expect?

\vskip-6pt
\vskip-6pt
\subsection{Our Block Cipher}
\vskip-6pt

Our examples are constructed for T-310, and old Feistel cipher with 4 branches.
%%% full version maybe %%%
%and one of the most important block cipher of the Cold War with some 3,800 cipher
%machines in active service in 1989 \cite{FeistCommunist,T-310,MasterPaperT310,T-310An80}.
This cipher offers great {\bf flexibility} in the choice of the internal wiring. %and
%so that we can possibly make some adjustments if we do not find a property we are looking for.
Most ciphers such as DES or AES also have this sort of flexibility in the choice of P-boxes,
arbitrary invertible matrices inside the S-box, inside the mixing layers, however later these components
are fixed. In T-310 this flexibility is ``officially'' supported:
a large variety of possible choices of cipher wiring can be specified and used.
Here if we find a weak setup, it will be entirely compatible with original historical hardware.
The exact cipher wiring specification in T-310 is called LZS or {\em Langzeitschl\"{u}ssel} cf.
\cite{MasterPaperT310} and various key studied by researchers
and various known complete specifications are denoted by 2 digit or 3 digit numbers such as LZS 31 or LZS 903,
cf. \cite{T-310Keys,MasterPaperT310}.
Our cipher uses Boolean functions on 6 variables which in our work will become a variable $Z$:
initially we study degenerated cases, which eventually at the end become a scenario where this Boolean function
is no longer chosen by the attacker, and a single attack works for a large number of possible Boolean functions.

\vskip-8pt
\vskip-8pt
\begin{figure}
\vskip-7pt
\begin{center}
\includegraphics*[width=5.1in,height=3.0in]{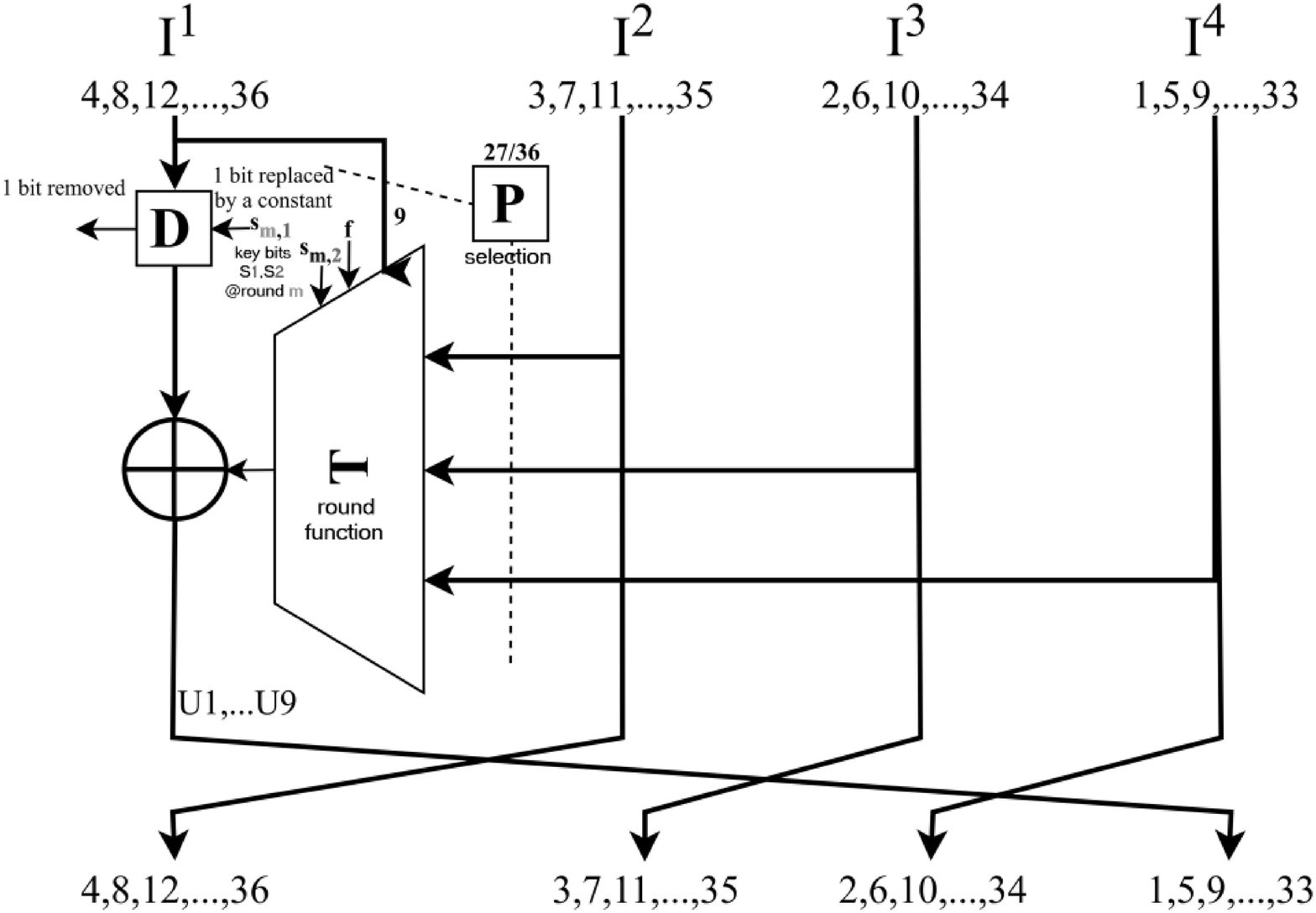}
\end{center}
\vskip-7pt
\vskip-7pt
\caption{T-310: a peculiar sort of Compressing Unbalanced Feistel scheme. %cf. \cite{UnbContractPata,MasterPaperT310}.
%In the common KT1 case,
%the spec allows to use also bits from the leftmost branch $I^1$
%under a number of highly technical conditions.
%It also disconnects$^{\ref{footnotedisconnect}}$
%ONE of the 9 bits in the left branch
%and replaces it by a key-dependent constant $s_{m,1}$ which is different in each round.
}
\label{FigContracting310KT1}
%\vskip-7pt
%\vskip-7pt
\end{figure}
\vskip-1pt

%\newpage

\vskip-4pt
\vskip-4pt
\subsection{Why Boolean Polynomials}
\vskip-6pt

An interesting question is why do we use Boolean polynomials in cryptanalysis.
Why not for example using roots of unity with $p=23$ cf. Section \ref{LameKummer}.
Potentially it is a natural and the best choice, and potentially there is no reason to do so.
The answer is that the arithmetic modulo 2 is {\bf a choice} of the attacker,
and potentially it is an arbitrary choice, and if it increases his chances of success,
another choice could be made.
A recent work done by a student \cite{MariosMSc} contains some highly detailed examples of how to construct
a simple non-linear invariant attack on a block cipher in several elementary steps.

\vskip-6pt
\vskip-6pt
\begin{figure}
\vskip-6pt
\hskip-10pt
\hskip-10pt
\begin{center}
\hskip-10pt
\hskip-10pt
\vskip-6pt
\vskip-6pt
\includegraphics*[width=5.1in,height=1.2in]{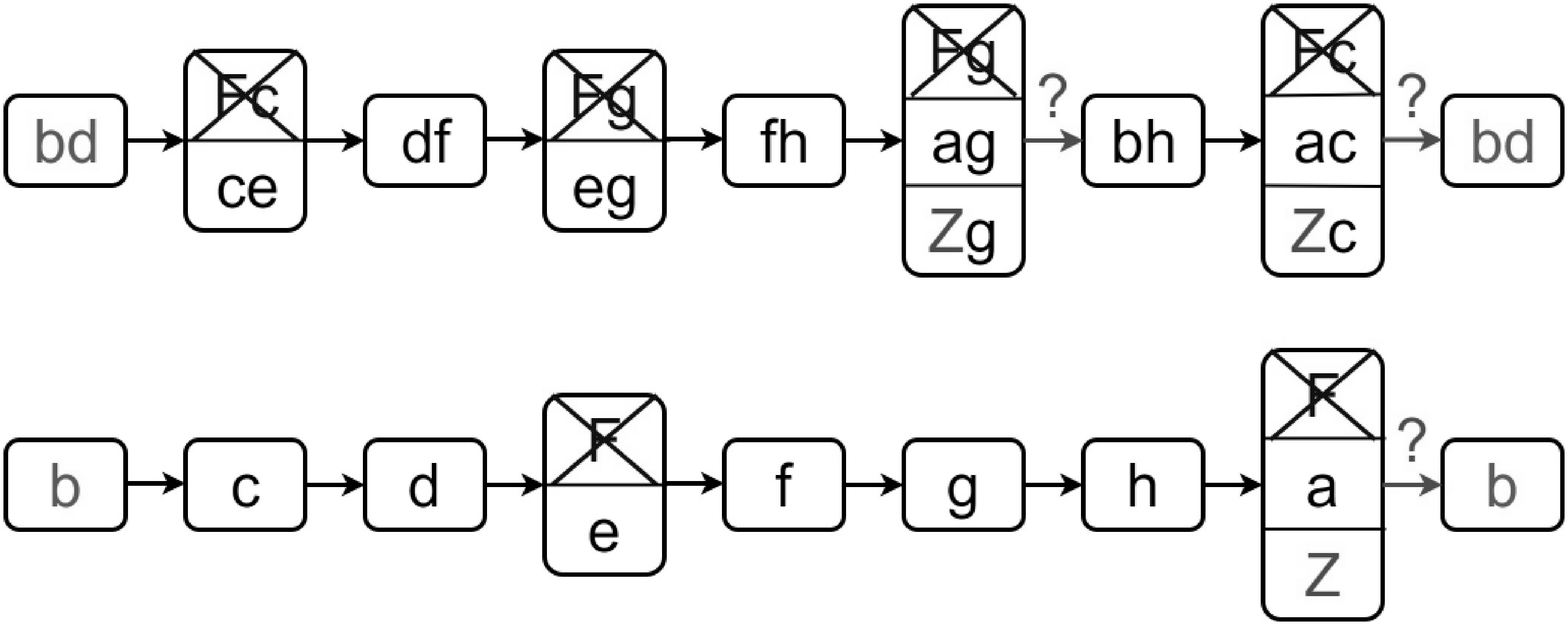}
\end{center}
\vskip-9pt
\vskip-9pt
\caption{A detailed example explaining how a certain invariant works \cite{MariosMSc}.}
%borrowed from \cite{BackdTut,MariosMSc}.}
\label{Fig2cycle827Z1}
\end{figure}
\vskip-9pt
\vskip-9pt

For example the polynomial $bd$ at the input is equal to the sum of term $Fc$ and polynomial $ce$ at the output.
We keep this term $Fc$ on the side hoping it will be eliminated later and focus on transitions of type
$bd\to ce$ for several steps. Then we form cycles where from $bd$ we come back to $bd$ as shown on
Fig. \ref{Fig2cycle827Z1} and in \cite{MariosMSc}.
At the end we sum all the monomials in black (or green) in one or several cycles
and hope that other terms such as $Fc$ in blue on Fig. \ref{Fig2cycle827Z1} below will appear an even number of times
and will therefore eventually be cancelled when we sum all the non-linear I/O equations together. %numerous monomials.
This is how we can show that the polynomial

\vskip-8pt
\vskip-8pt
$$
{\cal P}=a+b+c+ac+d+bd+e+ce+f+df+g+ag+eg+h+bh+fh
$$
\vskip-4pt
%160x0 has 4 x abcde=1 expected 5
%160x0 has 0 x abcdef=1 expected 2
%160x0 96x1
%also part of deg 2 is irreducible
%ax64 547777 a+b+c+ac+d+bd+e+ce+f+df+g+ag+eg+h+bh+fh
is an invariant for 1 round of encryption for a certain cipher setup  827, see Section 5.1 in \cite{ConstrRowEchKT1} and  \cite{MariosMSc}
for a more detailed explanation.

\vskip-4pt
\vskip-4pt
\subsection*{Attacks Modulo 3 or 4}
\vskip-6pt

From here it is easy to see that there is {\bf no obstacle} to construct an invariant attack modulo 3,
we just need to make sure that a number of variables or products we do not want to see
(like the in blue colour) in the final invariant,
will be divisible by 3 not by 2 and cancelled likewise.
This is maybe harder but not impossible.

Furthermore if we read a more recent paper \cite{BackdAnn} we are going to realize that in order to make
this sort of attacks more successful we need to maximize the chances that certain polynomials can be annihilated.
This suggests that maybe we need to look for attacks modulo 4, where $2\cdot 2=0$ could create additional opportunities
for annihilation.
Furthermore, we might need to follow the example of mathematicians (who have eventually proven the Fermat's last
theorem) and consider invariant properties in other rings or fields.
%The whole cipher cannot be made to work with same peculiar form of hidden algebraic low,
%but an invariant property for some bits could.
%further algebraic extensions of the rings in which cryptanalysis is typically done.
The choice of the arithmetic used in an attack ultimately lies with the attacker.

\vskip-6pt
\vskip-6pt
\subsection{Boolean Polynomials, Annihilators and Absorbers}
\vskip-6pt

Let $B_n$ be the ring of Boolean polynomials in $n$ variables (polynomials in their ANF without powers or with $x^2=x$ cancellations
done when multiplying the polynomials).
In this paper we do not use the annihilator method of \cite{BackdAnn} but we rather work on absorption properties,
a polynomial $f$ absorbs $g$ if $fg=f$.
In theory both sorts of event are equivalent, absorption of $g$ is the same as annihilation with $f(g+1)=0$.
However in practice in our attack we emphasise absorption and lack of unique factorization,
as key mathematical events which occur in order to make our attack work.

\vskip-6pt
\vskip-6pt
\subsection{On Mathematical Theory of Invariants.}
\vskip-6pt

%In mathematics
There exists an extensive theory of multivariate polynomial algebraic invariants %\cite{WolframInvariants}
%they were first called hyper-determinants by Cayley in 1845-1861, cf. \cite{CrillyInvariantsHist,CalayInvariants1845}.
%all modern terms in the theory of invariants such as invariant; covariant; comitant; discriminant, etc., were introduced by J. Sylvester.
w.r.t linear transformations going back to 1845 \cite{CalayInvariants1845,CrillyInvariantsHist,DieuCarrInvaTheo}. %,CalayInvariants1845}.
%%%while in the Soviet Union... \cite{GurevichInvariants}.........
This classical 19-th century invariant theory however deals with invariants
in the situations where (simultaneously):

\vskip-8pt
\vskip-8pt
\begin{enumerate}
\item [1)] invariants are polynomials of small degree,
\item [2)] they have only 2 sometimes up to 5 variables,
\item [3)] polynomials are over large fields and rings,
frequently algebraically closed or infinite (or both), or in fields with large characteristic,
\item [4)] invariants should not change when we operate a {\bf LINEAR} input variable transformation $L$,
a very important limitation,
\item [3+4]
makes that there is a scaling scalar or factor $\sigma$ in most invariants known in classical mathematics:
a determinant of the linear transformation $L$,
%cf Hilbert page 17, Cambridge re-edition 1993
\item [5)] these invariants are in general multivariate polynomials.
\end{enumerate}
\vskip-3pt

In modern invariant theory, however, there are of course more possibilities \cite{DieuCarrInvaTheo},
and here is what the first author of this book has put in a preface of his another (unrelated) book, cf.
slide 28 in \cite{CourtoisNewFroInvInd2008}:

\vskip-6pt
\vskip-6pt
\begin{quote}
[...] Everybody in mathematics knows that going
from one to several variables is an
important jump that is accompanied by great
difficulties and calls for completely new
methods [...]
\end{quote}
%[French Mathematician]
%Book "Calcul infinitésimal", Hermann, 1980
\vskip-3pt

In general however we are quite far from the traditional preoccupations of mathematicians.
%This is not at all what we do here!
The common points are:

\vskip-7pt
\vskip-7pt
\begin{enumerate}
\item [1)] we study polynomial invariants ${\cal P}$ of limited degree and
\item [5)] our invariants are multivariate polynomials over some fields,
\end{enumerate}
\vskip-3pt

However there are {\bf very substantial} differences:

\vskip-7pt
\vskip-7pt
\begin{enumerate}
\item [2')] we work with many more variables for ${\cal P}$, typically between 8 and 36 at a time.
\item [3')] we work on $GF(2)$ mainly,
\item [4')] and finally we are looking for %very different sorts of
invariants which remain the same after applying an extremely complex {\bf NON-LINEAR}
transformation called $\phi$, or any power of it $\phi^k$,
which are no longer linear cf. point 4) above,
very peculiar, %special, %%said above!!! and highly complex
%yet made to be somewhat implemented in hardware,
and not of the sort the mathematicians would consider worth studying,
%see Hilbert lecture LI 06 Aug 1897 page 187
\item [3+4]
Here the scaling factor $\delta$ could only be equal to $1$ and should be omitted.
\end{enumerate}
\vskip-4pt

\vskip-7pt
\vskip-7pt
\subsection{Round Invariants for Block Ciphers}
\vskip-6pt

A major risk in mathematics is that mathematical theories operate mainly at a syntaxic level
and they could be dealing essentially with an empty set.
This unless the objective is to prove the security by proving that the set is empty,
see \cite{FiliolNotVuln,BeiCantResNL}. % and later Section \ref{SecProveSecurityNoSols}.
Current research in application of polynomial invariants
in symmetric cryptography has lacked substance or material to work in the form
of real-life positive examples which work.
Numerous results are about cipher components rather than full ciphers.
For example for the AES-like S-box, it is possible to us the so called cross-ratio, %which is studied in
(which is already an invariant in the more general non-linear $\phi$ case
which is more rarely studied in mathematics).
% non-linear transformations
%
%Glynn (1998) has found the only known multiplicative hyperdeterminant in dimension larger than two.
%source: http://mathworld.wolfram.com/Hyperdeterminant.html
However this type of invariant is however still quite simple or we work with only one variable.
In our research we study a substantially wider variety of
multivariate invariants with increasing size and complexity.

In cryptographic invariants the main object to study are {\bf round invariants}
for one round. We would like to have
${\cal P}(\mbox{Inputs}) = {\cal P}(\mbox{Output ANF})$ where ${\cal P}$ will be a Boolean function.
We are looking for polynomials ${\cal P}$ the value of which does not change
after we apply a transformation called ``a round'' we call $\phi$.

This round function $\phi$ is typically a bijection and is like one round of encryption.
In addition, typically it is NOT one fixed permutation but it has a parameter,
a secret key and potentially additional parameters.
The more parameters, the harder it becomes to find invariants.
For example the T-310 cipher can be viewed as each round is applying one of the 8 possible permutations
$\phi_0: \{0,1\}^{36}\to \{0,1\}^{36}$ up to $\phi_7: \{0,1\}^{36}\to \{0,1\}^{36}$ and the choice which $\phi$ is actually used depends
on 2 bits of the secret key and 1 bit of the IV (which is public and known to the attacker),
all using the original notations of \cite{T-310An80}.
Technically speaking, just finding such invariants is easy and they exist in vast numbers
yet many are in some sense trivial or degenerated, cf. \cite{BackdTut}.
A key problem is to find {\bf simultaneous invariants} to hold in all the eight cases,
i.e. for all of $\phi_{0}\ldots \phi_{7}$ simultaneously.
%cf. later Sections
%\ref{CanWeHaveTheSameInvariantForF0F1} and \ref{ConstructSimultF0F1},
%and such invariants should occur without being a consequence of simpler linear invariants,
%cf. Appendix  \ref{FEReductionToZeroExample3}.
This was a big problem in early research on this topic cf. \cite{BackdTut}
but it is not a problem in this paper (we construct a solution which works
directly and there are many more variables which are eliminated also). %(our methodology allows .

\vskip-8pt
\vskip-8pt
\subsection{Group Theory vs. Invariants for Block Ciphers}
\vskip-6pt

We recall that for the finite field ``inverse'' S-box, and NOT for the actual AES S-box, cf. \cite{invglc},
it is possible to use the so called cross-ratio.  %which is studied in
Again this is already an invariant in our more general non-linear $\phi$ case.
We refer to Section 4 in \cite{invglc} for more details and further references.
%This cross ratio invariant is NOT always correctly applied.
%In fact there are singularities in polynomial invariants, and
%the cross-ratio invariant is simply not always true, sometimes it ``breaks'',
%or there is a discontinuity cf. \cite{invglc}, this correspond to invariants being correct
%probability of type $1-\varepsilon$ with some small $\varepsilon$.
%%
%This fact plays an important role inside the so called
%%where it is shown that for the inverse function used in the AES S-box, the cross-ratio is not an invariant,
%%but rather an almost-invariant with a potential discontinuity which occurs with a low probability.
%%This leads to
%``Whitening Paradox'' \cite{invglc,WhiteningParadox}
%which is about security (and insecurity) of block ciphers and it is highly relevant here.
%The essence of why we have a paradox here can be briefly summarized as follows.
The ``Whitening Paradox'' paradox is a proof of concept that a group-theoretic claims in cryptography
\cite{T-310An80,DESPrimitiveWernsdorf,AESWernsdorf,SalaGOSTgroup} can be highly misleading
and can lead to a ciphers where the group of transformations generated by the cipher is
proven mathematically to be extremely large,
and which are nevertheless insecure and can simultaneously broken for an exponentially large number of rounds.
%This even though it is clear that when the number of rounds grows further,
%they are ultimately secure, cf. \cite{invglc,WhiteningParadox}.
%In other words we get a cipher which is simultaneously provably secure and practically insecure.
We refer to \cite{invglc,WhiteningParadox} for more details.

There exists numerous modern works on the group of transformations generated by a block cipher
\cite{DESPrimitiveWernsdorf,KennyImprimitive,AESWernsdorf,SalaGOSTgroup,invglc}
and primitive groups,
\cite{DESPrimitiveWernsdorf,KennyImprimitive,SalaGOSTgroup,ItalianNLBackdoorsBlock}.
This research topic was very clearly was invented during the Cold War
and was already studied very carefully in the 1970s
with very specific security claims which are contained in \cite{T-310An80},
however these claims are not formulated as precise mathematical theorems in \cite{T-310An80}
and are therefore subject to interpretation.

\vskip-6pt
\vskip-6pt
\subsection{Weak Ciphers and Bacdoors}
\vskip-6pt

A first impression when reading this paper is that it is
our paper is about how to ``backdoor'' a block cipher:
how to make it weak on purpose \cite{invglc,WhiteningParadox,KennyImprimitive,WhatDESay,gostdc2bHistGoodSboxes,ItalianNLBackdoorsBlock,BannierPartBack,MaliciousSHA,MaliciousKeccak}.
This first impression is rather totally incorrect:
what we do rather ``proper'' cryptanalysis of block ciphers.
This is because absolutely every property\footnote{Such as assumptions on $P$ and $D$,
or the fact that the Boolean function will have a cubic annihilator or absorber of some peculiar form.}
we study about the cipher wiring happens with a relatively large probability.
Overall we will get an attack which may happen accidentally, also when the whole cipher specification
and the non-linear components are the strongest possible and were not chosen by the attacker.
Moreover a wider variety of such attacks exist.
If our polynomial invariant ${\cal P}$ of degree 7 does not work, another similar property could eventually work
%Then if our polynomial invariant ${\cal P}$ of degree 7 does not work, another one could work.
and the success probability is likely to be higher than it seems from the strict attack presented here below.

%\vskip-6pt
%\vskip-6pt
\subsection{On Irreducible Polynomials}
\label{IrreducibleRebuttal}
\vskip-6pt

Recent papers show how to construct polynomial non-linear invariant attacks on some block ciphers.
Some such attacks are clearly trivial, for example products of linear polynomials
which are already invariants for the same block cipher\footnote{
For example in appendix of \cite{BackdTut} the author found an invariant with $169=13^2$ terms
which %was a bad sign already
and it turns out to be equal to a
%$
%(n+b+p+r+t+v+x+z+N+P+R+T+V)%\cdot
%(a+m+o+q+s+u+w+y+M+O+Q+S+U)
%$
a product or two linear polynomials with 13 terms which are also round invariants.
Such examples contain nothing new w.r.t. Matsui's Linear Cryptanalysis.
}.
No new attack is found in this case.
For this reason some early papers on this topic emphasise
invariants being irreducible polynomials \cite{BackdTut,ConstrRowEchKT1}.
%For example an invariant property which works for a large number
%of rounds such being a prime as 7 or 127 is likely to be of interest,
%or an irreducible polynomial of degree being also a prime.
%Another question we study is whether it is possible to find correlation attacks which appear ``ex-nihilo'' i.e.
%they are NOT at all a combination of some simpler correlations such as in Linear Cryptanalysis.

Irreducible polynomials turns out to be a false good idea.
In general there is no problem whatsoever with products of linear polynomials:
in general these linear factors will NOT correspond to any attack on the same cipher
(and can only correspond to an attack a substantially weaker cipher).
This is very clearly demonstrated in recent papers \cite{BackdTut,InvHopWCC} and the strongest
round invariant attacks ever found are of this type \cite{BackdAnn}.

\vskip-6pt
\vskip-6pt
\subsection{On Sporadic Attacks vs. Product Construction}
\vskip-6pt

An interesting question is discovering some ``sporadic'' %and more ``primary''
properties the existence of which we have not anticipated. %which are not easily found and constructed.
The attack presented was actually first discovered accidentally,
by trying the exact cipher setup 265 of \cite{BackdAnn} with various Boolean functions chosen at random.
We discovered that they occur.
Only later we developed a detailed mathematical explanation about WHY such an attack may work.
At several we find it quite surprising that such an attack may at all be ever made to work on any block cipher.
Indeed we make two extremely complex polynomials in 16 variables to be equal for any input on 36 bits,
for any key, any IV (and any number of rounds in further cryptanalytic applications).
Moreover the are 7 polynomials and there no trivial transitions of type 4+4 due to the Feistel structure
with 4 branches which was a key feature in previous attacks \cite{BackdAnn,MariosMSc}.

After discovering this new invariant we realized that the polynomial can be factored.
%(factorization is not unique)
We belive that this is NOT accidental.
Not only a strong attack in \cite{BackdAnn} is a product of linear polynomials
but also the new unexpected attack is such.
This is really one of the main claims of this paper:
the product construction of \cite{BackdAnn} and another new and (substantially more complex) product construction
in this paper are not a good attacks by accident.
They correspond to a powerful method of constructing attacks and there is some sort of phase transition
where attacks become easier to construct as the degree of this product increases.
This is because higher degree monomials can be obtained in several different ways,
hence opportunities for cancelations, or obtaining the same polynomial in different ways.
In general well-made polynomials which are products of many terms are MORE likely to lead
to the sort of cancellations we need in order to make an attack work.
All this is conjectured but not proven (better attacks could also exist elsewhere)
and in our attack we need to make two polynomials being products of many factors equal, not just annihilate one such product,
which however seems to be achievable roughly for the same reasons.

%\subsubsection{Artificial Versus Real Scarcity.}
%Most block cipher attacks ever found in block ciphers operate through combination
%of simpler attacks. In this paper we show that other sorts of attacks exist
%and possibly that we need to considerably enlarge our bag of tools and tricks in cryptanalysis or we are going
%to miss a lot of attacks.
%%if we just continue incremental combinations of known attacks.
%The real scarcity of new attacks could be a combinatorial one, in the sense
%that trivial attacks are more numerous and may hide the existence of more sporadic attacks
%of less trivial types.

%\newpage

\vskip-6pt
\vskip-6pt
\section{Linear Cryptanalysis - New Possibilities}
\label{IntroLin127}
\vskip-6pt

Linear cryptanalysis is a popular research topic since Matsui, \cite{Matsui}.
However recent research shows that it is older than initially thought
%it looks like it
%in Eastern Bloc the study of linea was a routine technique for studying ciphers and their components already in the 1970s i
cf. \cite{LCKT1ucry18}. There exist numerous papers on linear attacks and such attacks are a tough game.
They require large quantities of data encrypted with a single key and properties are highly regular.
For example the best attack on DES by Matsui depends on a property with a period of 14 rounds.
For T-310 there exist numerous examples where the periodic property has say 6 or 8 or 12 rounds,
and extremely few for say 13 rounds, cf. \cite{MasterPaperT310}.

At the same time if we read the specification of T-310 cf. \cite{T-310},
we will see that state bits which are actually
used for encryption will be extracted at the speed of 1 bit every 127 rounds, where 127 is a prime.
We also learn that key bits are repeated every 120 rounds.
Finally we learn that the IV bits are repeated every $2^{61}-1$ bits
which is also a prime, and this choice was clearly
a deliberate choice by the designers, cf. \cite{T-310An80}.
So there is no hope that we could find any ``relevant'' periodic invariant attack, right,
and for example it is unthinkable that an invariant with a period being a larger prime such as 127 would exist 127
for a block cipher?
In this paper we show that an invariant with a period of 127 may exist for a block cipher.
Moreover such invariants do exist for some real-life keys.
Such a degree of complexity was never seen before in Linear Cryptanalysis.

\vskip-6pt
\vskip-6pt
\subsection{A New Discovery - a Large Prime Periodic Property}
\vskip-6pt

We found that when we set the Boolean function to zero (equal to zero for every input)
the real-life setup LZS 31 and LZS 33 of cipher T-310 exhibits a
complex linear property valid for
127 rounds.
We should insist on the fact that 127 is a prime,
and this is a solid irreducible complex invariant property
which is most likely not a consequence or a combination
of any simpler periodic properties.
The number of active bits (out of 26) follows a complex pattern:

\begin{verbatim}
12->12->14->16->15->15->17->17->16->18->16->17->18->17->19->18->15->16->15->
13->14->16->18->20->23->22->22->22->21->20->20->21->23->23->22->21->21->19->
and back to 12
\end{verbatim}

\noindent
%The number 127 is highly relevant to the periodicity inside the encryption, however
Is there any hope that this property could be ever used in cryptanalysis?

%\vskip-2pt
%\vskip-2pt
%\begin{verbatim}
%LZS31 /Bool0
%LZS33 /Bool0
%LZS133 /Bool0
%LZS127 /Boole
%the last one has two disjoint cycles of len 127
%\end{verbatim}
%\vskip-3pt
%%systematic: /Bool0 or /Bool1, both work, same with e and e+1
%%127: P=4,2,33,12,32,16,5,15,9,10,24,35,36,14,21,11,1,
%%25,8,28,22,20,3,29,30,34,6 D=0,24,4,20,12,32,36,16,8
%%LZS127 /Boole+1 also
%%LZS133 also
%%and no other as it seems, but my code is NOT so reliable for 127R

\vskip-6pt
\vskip-6pt
\subsection{How to Maybe Break T-310 in a Real-Life Historical Setting}
\vskip-6pt

In this section we sketch how this property can maybe be used to decrypt T-310 communications.
The attack is hypothetical and is meant to highlight the role of algebra in constructing
an attack of the sort never seen before. A possible attack will work as follows:

\begin{enumerate}
\item
We consider one of the real-life cipher variants LZS 31 or 33 cf. \cite{T-310Keys,MasterPaperT310}.
\item
We express the problem of finding an invariant for 127 rounds as a system of algebraic equations
where the unknowns are the coefficients of the Algebraic Normal Form (ANF) of the Boolean function $Z$.
We call $FE$ this equation, cf.
which is a simple I/O sum of two polynomials (an Input polynomial and an Output polynomial)
cf. %and is formally defined much later in
Section \ref{FEdef}.
\item
In general such equation has no solutions and it is very hard to know if it has any.
However here we already have a case where $FE$ has a solution!
\item
Now we multiply our 127 polynomials by a well-chosen polynomial at each step.
We get a more complex $FE$ which is expected to have more than one solution
(if we are lucky or if the polynomials were well chosen).
%\item
%A major variants is to reduce the number of steps: construct a complex invariant for one round.
\item
At the end if we do it well, this could work for a Boolean function not chosen by the attacker.
The key property is that there are many ways to annihilate Boolean functions in this type of attacks
which are sufficiently powerful in order to attempt to break ciphers with arbitrary Boolean functions,
see \cite{BackdAnn} and this paper for specific examples.
\item
At the end we might be able to construct an invariant attack with an invariant with a prime period of 127
or with another completely different prime period.
\item
A period of 127, given that bits used for encryption also obey a period of 127,
should lead to powerful attacks which allow to decrypt communications.
\end{enumerate}

We have not yet presented any non-linear attack yet,
however we have a powerful framework or methodology to construct one.
From a trivial completely linear attack initially, or/and from set of polynomials which represent a hypothetical attack which does not work
or where the Boolean function is degenerated, we can attempt to construct an attack which works
for more complex random Boolean functions.
If the reader doubts whether this methodology works we refer to \cite{InvHopWCC} which contains
an elaborate complete proof of concept where the cipher is modified several times simpler linear attacks are removed
and complex high degree invariants only are kept, while the complexity of the invariants and
the complexity of the Boolean function also increases.

\vskip-6pt
\vskip-6pt
\subsubsection{What is Next?}
\vskip-6pt

In what follows we will construct a new elaborate example of an invariant polynomial of degree 7.
% which shows that this sort of methodology can be made to work.
We will however construct an invariant for one round (instead of 127 rounds).
In fact it is potentially extremely difficult to work with 127 rounds directly in the general non-linear case.
The size of the polynomial equations to study and the size of $FE$ to solve would literally explode.
Substantial simplifications are needed (or some way to limit the size of polynomials)
and we are still learning how to build non-liner attacks on block cipher with our new white-box methodology.
Our ultimate goal would be to obtain something which works for a non-negligible fraction such as say 1 $\%$
of all Boolean functions on 6 variables, cf. \cite{BackdAnn}, which goal will be achieved later inside this paper.
In all cases an important hint is that invariants where the polynomial is a product of several simpler polynomials
are quite powerful %and can break block ciphers,
cf. Section \ref{IrreducibleRebuttal} above and \cite{BackdAnn}.
%As another new proof of concept we are now going to construct attack where 7 different polynomials will multiplied together.

\newpage

\vskip-6pt
\vskip-6pt
\section{
Non-Linear Cryptanalysis
through Formal ANF Coding
%through Polynomial Coding
%%not at all here yet just coding ANFs and Constructing
%Polynomial Invariant Attacks
}% on Block Ciphers}
\vskip-5pt

The concept of cryptanalysis with non-linear polynomials
a.k.a. Generalized Linear Cryptanalysis (GLC)
was introduced
%earlier
%by Harpes, Kramer, and Massey
at Eurocrypt'95, cf. \cite{GenLinear1}.
A key question is the existence of round-invariant I/O sums: %or I/O relations
when a value of a certain polynomial is preserved
%equation, written separately for the Input and Output,
%which are invariant
after 1 round.
%cf. \cite{SlidesDESteach}.
%Finding %Constructing
%such properties is very difficult %combinatorial problem,
Many researchers have in the past failed to find any such properties,
%cf. for example Knudsen and Robshaw at Eurocrypt'96 cf. \cite{GenLinear2}. %for major ciphers.
Bi-Linear and Multi-Linear attacks were %subsequently
introduced \cite{BLC,invglc} %,SlidesDESteach}
%section B.2., D.5 and H of invglc extended eprint version!
%in order to work with
for Feistel ciphers %with two and several
branches specifically.
%The number of such attacks grows as $2^{2^n}$, many authors
%stress that systematic exploration is not feasible \cite{BeiCantResNL}.
%For this reason, we there are extremely few positive results on this topic \cite{TodoNL18,BackdTut}
%and any method to approximate the solution is valuable,
%as one working example as in this paper will typically allow the researchers
%to find more similar examples.
In this paper and unlike in \cite{TodoNL18} we focus on invariants which work for 100 $\%$
of the keys and we focus on stronger invariants which hold with probability 1.
%In addition we look at a strong us case: an expensive government cipher in which the
%block cipher is used for encryption in an extremely low-data rate mode,
%a lot more costly than 3DES, AES, cf. \cite{FeistCommunist}.
%Here most cryptanalytic attacks simply do not work (!).
%%Moreover, even the existence of a number of linear higher degree invariant does not seem an immediate threat.
%However all this complexity is not that useful if we are able to construct
%powerful non-linear invariants which work for any number of rounds.

We call ${\cal P}$ a polynomial invariant if the value of ${\cal P}$
is preserved after one round of encryption,
i.e. if ${\cal P}(\mbox{Inputs})={\cal P}(\mbox{Output ANF})$.
In this paper we work with one specific block cipher with 36-bit blocks.
%a proof of concept that non-linear cryptanalysis can be made to work.
%We do not provide a full description of an encryption system
%(how it is initialized and used etc).
The main point is that any block cipher round translates into relatively simple Boolean polynomials,
if we look at just one round.
We follow the methodology of \cite{BackdTut} in order
to specify the exact mathematical constraint, known as the Fundamental Equation or $FE$,
cf. Section \ref{FEdef},
so that we could have a polynomial invariant attack on our cipher.
Such an attack will propagate for any number of rounds (if independent of key and other bits).
In addition it makes sense following \cite{BackdTut} to consider
%%% full version maybe %%%
%which was rarely done in cryptanalysis,
that the Boolean function is an unknown. % yet to be determined.
We denote this function by a special variable $Z$.
We then see that our attack works if and only if $Z$ is a solution to
a certain algebraic equation [with additional variables].
%Then and quite surprisingly, sometimes, $FE$ reduces to zero,
%and the attack works for any Boolean function.
%This research is full of good surprises,
%there are numerous things which work better than expected.
The main interest of making $Z$ a variable is to see that
even if $Z$ is extremely strong, some advanced ``product''
attacks will work nevertheless.

%\vskip-8pt
%\vskip-8pt
\subsection{Notation and Methodology}
\vskip-5pt

%In order to have notations, which are as compact as possible,
In this paper the sign + denotes addition modulo 2, and frequently
we omit the sign * %or $\cdot$
in products.
For the sake of compact notation we frequently use short or single letter variable names.
For example let $x_{1},\ldots, x_{36}$ be inputs of a block cipher each being $\in\{0,1\}$.
We will avoid this notation and name them with small letters $a-z$ and letters $M-V$ when we run out of lowercase letters.
%and  or $e_1$ for various binary variables $\in\{0,1\}$.
%36+'a'-'c'=34=c
%30=g 28=i 26=k 24=m 22=o
We follow the backwards numbering convention of \cite{BackdTut} with $a=x_{36}$ till $z=x_{11}$
and then we use specific capital letters $M=x_{10}$ till $V=x_{1}$.
This avoids some ``special'' capital letters
%names $S1,S2,K,L,F$ are used as placeholders for something more complex
following notations used since the 1970s \cite{FeistCommunist,T-310,T-310An80}.
We consider that each round of encryption is identical except that
they can differ only in some ``public'' bits called $F$,
a round constant\footnote{It is different in each round
and it also known as IV bits
which are derived from na LFSR with a very large period cf. \cite{T-310}.}.
known to the attacker
and some ``secret'' or key bits called $S1$ or $K$ and $S2=L$.
%We call $S1=K$ and $S2=L$ two key bits used in one round, and $F$ is a round constant (public) bit,
Even though these bits ARE different in different rounds we will omit to specify in which round
we take them because our work is about constructing {\bf one round} invariants (extending to any number of rounds).
This framework covers most block ciphers ever made  % and ever studied in human history.
except that some ciphers would have more ``secret'' or ``public'' bits in one round.
The capital letter $Z$ is a placeholder for substitution of the following kind

\vskip-4pt
\vskip-4pt
$$
Z(e_1,e_2,e_3,e_4,e_5,e_6)
$$
\vskip-1pt

where $e_1\ldots e_6$ will be some 6 of the other variables.
In practice, the $e_i$ will represent a specific subset of variables
of type $a$-$z$, or other such as $L$.
At the end $Z$ must be replaced by a formula like:

\vskip-4pt
\vskip-4pt
$$
Z \leftarrow Z00+Z01*L+Z02*c+Z03*Lc+\ldots +Z62*cklfh+Z63*Lcklfh
$$
\vskip-1pt

where $Zij$ are coefficients of the Algebraic Normal Form (ANF).

\vskip-8pt
\vskip-8pt
\subsection{Constructive Approach Given the Cipher Wiring}
\vskip-6pt

Our attack methodology starts
%\footnote{
%Our approach is to find invariant attack starting from arbitrary rounds ANFs
%is at the antipodes compared to \cite{invglc,WhiteningParadox} where the ciphers are very special.
%% and have very strong %not hidden!!!
%%high-level structure.
%}
 from a given block cipher specified by its ANFs for one round.
 %%(possibly with some ``chose permutation of wires'' flexibility for connecting non-linear components)
%and we search for complex polynomial invariant properties.
%Our approach is applicable to more or less
%%\footnote{
%%This is if there is a sufficient amount of entropy inside these non-linear components,
%%possibly not to Simon or TEA where S-boxes are too small or rather inexistent
%%for us to manipulate.}
%any block cipher which contains non-linear components,
%and also to most hash functions and stream ciphers
%based on a core block cipher.
%% (a large permutation
%%with few round-dependent bits which can be key bits, IV bits
%%or message bits). % construction.
%A success of this approach is measured by key space ratios:
%how many long-term keys are affected by the attack.
%If this ratio is high for example $10 \%$, and if 10 different
%years 10 different long-term keys are used,
%we can hope to break sooner or later and if we are lucky,
%maybe one of the real-life long-term keys.
%Too many block ciphers which appear in cryptographic literature are never used in any real-life situation.
Specific examples will be shown for T-310.
The block size is 36 bits and the key has 240 bits.
%intro done earlier
%, and old Feistel cipher with 4 branches
%and undoubtedly the most important block cipher of the Cold War with some 3,800 cipher
%machines in active service in 1989 \cite{FeistCommunist,T-310,MasterPaperT310,T-310An80}.
%%This cipher offers great {\bf flexibility} in the choice of the internal wiring and
%%entirely compatible with original historical hardware.
The hardware %complexity and
encryption cost with T-310 is
hundreds of times bigger than AES or 3DES, cf. \cite{%LCKT1ucry18,
FeistCommunist}.
%However one round is {\bf intelligible} and by no means secure.
%Therefore the designers have mandated that incredibly large number of 1651 such rounds must be executed in order
%encrypt just one character on 5 bits.
Does it make this cipher very secure?
% as intended for a serious government security cipher in the middle of the Cold War?
Not quite, if we can %are able to
construct
algebraic invariants which work for any number of rounds.

\vskip-6pt
\vskip-6pt
\subsection{ANF Coding of One Full Round}
\label{GeneralANDSubstCoding}
\vskip-6pt

%%Let us imagine that we are presented with an arbitrary complex block cipher designed
%%by some very paranoid designers with large complexity and a sophisticated internal structure.
%In the picture below we show the internal structure of T-310.
%The cipher operates on 36-bit blocks and the
%In this %compressing unbalanced Feistel cipher with 4 branches
We number the cipher state bits from 1 to 36 where bits $1,5,9\ldots 33$ are those freshly created
in one round, cf. Fig \ref{FigContracting310KT1}.
%while ALL the input bits the numbers of which are NOT multiples of 4 are shifted by 1 position, i.e.
%bit 1 becomes 2 in the next round, and bit 35 becomes 36.
Let $x_{1},\ldots ,x_{36}$ be the inputs and
let $y_{1},\ldots ,y_{36}$ be the outputs.
One round of our cipher can be described as 36 Boolean polynomials out of which only 9 are non-trivial:

%written by WriteSimpleSubstitutionFile11to36 with a-z letters
%{\scriptsize
%\vskip-9pt
%\vskip-9pt
\begin{align*}
y_{33}  &= F + x_{D(9)}  & \\
& Z1 \hspace{-.28em}\stackrel{def}{=}\hspace{-.28em}Z(S2,x_{P(1)},\hspace{-.20em}\ldots,\hspace{-.20em}x_{P(5))})\hspace{-.55em}\hspace{-.70em}\\
y_{29}  &=  F  + Z1 + x_{D(8)} & \\
y_{25}  &=  F  + Z1 + x_{P(6)}+x_{D(7)} & ~~~~~~~~~~~~~~~~~~~~~~~~~~~ &\\
& Z2 \stackrel{def}{=}Z(x_{P(7)},\ldots, x_{P(12)})\\
y_{21}  &=  F  + Z1 + x_{P(6)}+Z2+ & x_{D(6)} ~~~~~~~~~~~~~~~~~~~~~~~~~~~~& \\
y_{17}  &=  F  + Z1 + x_{P(6)}+Z2+ &  x_{P(13)} + x_{D(5)} ~~~~~~~~~~~~~~~~& \\
& Z3 \stackrel{def}{=}Z(x_{P(14)},\ldots, x_{P(19)})\\
y_{13}  &=  F  + Z1 + x_{P(6)}+Z2+ & x_{P(13)} + S2 + Z3 + x_{D(4)} ~& \\
y_{9}  &=  F  + Z1 + x_{P(6)}+Z2+ & x_{P(13)} + S2 + Z3 + x_{P(20)} &+x_{D(3)}~~~~~~~~~~~~~~~~~ \\
& Z4 \stackrel{def}{=}Z(x_{P(21)},\ldots, x_{P(26)})\\
y_{5}  &=  F  + Z1 + x_{P(6)}+Z2+ & x_{P(13)} + S2 + Z3 + x_{P(20)} &\hspace{-.15em}+\hspace{-.15em}Z4 \hspace{-.15em}+\hspace{-.15em} x_{D(2)}\hspace{-.15em}~~~~~~~~~~~~  \\
y_{1}  &=  F  + Z1 + x_{P(6)}+Z2+ & x_{P(13)} + S2 + Z3 + x_{P(20)} &\hspace{-.15em}+\hspace{-.15em}Z4 \hspace{-.15em}+\hspace{-.15em} x_{P(27)} \hspace{-.15em}+\hspace{-.15em} x_{D(1)} \\
 & x_{0} \stackrel{def}{=}S1 & \\%\mbox{~e.g. if~} D(1)=0 \mbox{~then $S1$ appears in the last equation}\\
y_{i+1}&=x_{i} \mbox{~for all other~} i\ne 4k & (\mbox{~with~} 1\leq i\leq 36)~~~~~~~~~~\\
\end{align*}
%\vskip-7pt
%\vskip-7pt
%}

%\newpage

Two things remain unspecified: % on our picture:
the $P$ and $D$ boxes or the internal wiring.
%which bits and in which order are connected to D1-D9 and v1-v27.
In T-310 this specification is called an
LZS or {\em Langzeitschl\"{u}ssel}
which means a long-term key setup.
%%% full version maybe %%%
%and which is distinct than the short-term key on 240 bits.
%It could be compared to the knowledge of rotors in Enigma or S-boxes in GOST.
%
We simply need to specify
two functions
%\footnote{
%Which are both assumed to be injective and $D(i)$ are frequently always multiples of $4$.
%%this in order to avoid many degenerate cases and trivial attacks cf. \cite{MasterPaperT310}.
%}
$D: \{1\ldots 9\} \to \{0\ldots 36\}$, $P:\{1\ldots 27\}\to \{1\ldots 36\}$.
For example $D(5)=36$ will mean that input bit 36 is connected
to the wire which becomes $U5=y_{17}$ after XOR of Fig. \ref{FigContracting310KT1}.
Then $P(1)=25$ will mean that input 25 is connected as v1 or the 2nd input of $Z1$.
We also apply a special convention where the bit S1 is used instead of one of the $D(i)$ by specifying that $D(i)=0$.
%%% full version maybe %%%
%The expressions become considerably simpler if we fix the long-term key LZS wiring
%as shown in Section \ref{SubsEqsShorter} below.

\vskip-5pt
\vskip-5pt
\begin{figure}
\vskip-5pt
\vskip-5pt
\hskip-10pt
\hskip-10pt
\begin{center}
\hskip-10pt
\hskip-10pt
\vskip-6pt
\vskip-6pt
\includegraphics*[width=5.1in,height=2.8in]{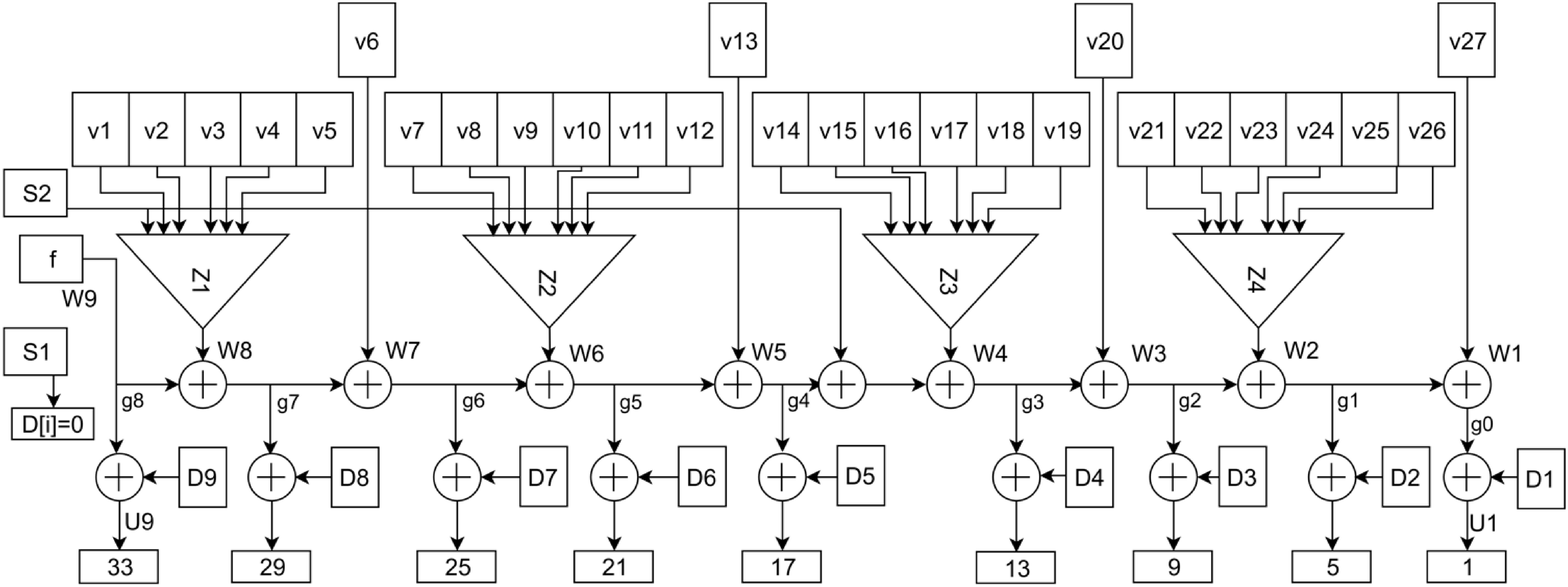}
\end{center}
\vskip-7pt
\vskip-7pt
\vskip-7pt
\caption{The internal structure of one round of T-310 block cipher.
}
%inside the ``complication unit''
%%see also page 119 of \cite{T-310An80}
%Two figures at Drobick page ke-sks-t310 clearly show that complication unity in T-310 has 1 less output 9 vs 10 for SKS
\label{FigComplicationUnit6basic}
\vskip-6pt
\vskip-6pt
\end{figure}
\vskip-4pt
\vskip-4pt

%\newpage
%
%\vskip-6pt
%\vskip-6pt
%\subsection{Internals of One Round}
%\vskip-6pt

%\newpage

\vskip-4pt
\vskip-4pt
\subsection{The Substitutions.}
\label{ExactSubstitutions2}
\vskip-4pt

Overall one round can be described as 36 Boolean polynomials
of degree 6; out of which only 9 are non-trivial.
One round of encryption is viewed as a sequence of substitutions
where an output variable is replaced
by a polynomial algebraic expression in the input variables.
%%%The full formulas in the general case can be found in \cite{BackdTut}.
Here is a (shortened) example following the cipher specification step-by-step
for the long-term key 551 used in \cite{BackdTut}:

%ax64 856317 /LZS551 P=abcdef
\vskip-9pt
\vskip-9pt
\begin{align*}
a  &\leftarrow  b & \\
b  &\leftarrow  c & \\
c  &\leftarrow  d & \\
d  &\leftarrow  F + i & \\
& [\ldots] & \\
%&Z1 \leftarrow Z(L,t,S,d,y,m))\\
%l  &\leftarrow  F + Z1 + O + q & \\
& [\ldots] & \\
%& Z4 \leftarrow Z(w,u,a,h,e,n)\\
%& [\ldots] & \\
V &\leftarrow F + Z1 + O + Z2 + q + L + Z3 + i + Z4 + k + K&\\
\end{align*}
\vskip-7pt
\vskip-7pt
%}
\label{SubsEqsShorter}
\vskip-7pt

%%% full version maybe %%%
%We emphasise the fact that the variables $x_i$ and $y_i$ are treated ``alike''
%are called be the SAME letter, for example $a=x_{36}$ and $a=y_{36}$, as we study round invariants.
%%% full version maybe %%%
%We recall that $F$ is a public bit derived from an IV transmitted in the cleartext,
%S1 and S2 are bits of the secret key on %which has
%240 bits. S1 and S2 are repeated every 120 steps.
In order to have shorter expressions to manipulate %later on,
we frequently replace $Z1-Z4$ by shorter abbreviations $Z,Y,X,W$ respectively.
%We also sometimes replace a single letter which comes from P6 by a variable $G$. %%P27 is J
We also replace S2 by a single letter $L$ (used at 2 places).
The other key bits $S1=K$ will only be used %in at most one place,
if some $D(i)=0$.

%\newpage

\vskip-6pt
\vskip-6pt
\section{The Fundamental Equation}
\label{FEdef}
\vskip-1pt

In order to break our cipher we need to find a polynomial
expression ${\cal P}$ say

\vskip-5pt
\vskip-5pt
$$
{\cal P}(a,b,c,d,e,f,g,h,\ldots) =
abcdijkl+efg+efh+egh+fgh
$$
\vskip-1pt

using any number between 1 and 36 variables
such that if we substitute in ${\cal P}$  all the variables by the substitutions defined
%above in Fig. \ref{SubsEqsShorter},
we would get exactly the same polynomial expression ${\cal P}$, i.e.
$
{\cal P}(\mbox{Inputs}) =
{\cal P}(\mbox{Output ANF})
$ are equal as multivariate polynomials.
%%%later
%The key observation is that we can actually {\bf guarantee} that ${\cal P}$ is an invariant
%attack on our cipher by solving this relatively simple algebraic equation
%in order to determine the unknown Boolean function $Z$ and the coefficients of the
%polynomial ${\cal P}$. Then we win if (in this paper we are quite lucky!)
For example (or to start with) we assume that this polynomial ${\cal P}$ is fixed.
Then the attacker will write ONE SINGLE algebraic equation which he is going to solve
to determine the unknown Boolean function $Z$, if it exists.
%\newpage
There are several forms of his equation written with more less laconic vs. very precise notations.

\vskip-6pt
\vskip-6pt
\subsection{Compact Notations - Basic High-Level Fundamental Equation}
%\vskip-1pt

\begin{defi}[Compact Uni/Quadri-variate FE]
\label{defiCompactFE}
Our ``Fundamental Equation (FE)'' to solve is to make sure that sum of two polynomials like:

\vskip-3pt
\vskip-3pt
$$
FE = {\cal P}(\mbox{Inputs}) + {\cal P}(\mbox{Output ANF})
$$
\vskip-2pt

reduces to 0, or more precisely we are aiming at $FE=0$ for any input,
or in other words we want to achieve a formal equality of two Boolean polynomials like

\vskip-3pt
\vskip-3pt
$$
{\cal P}(\mbox{Inputs}) +
{\cal P}(\mbox{Transformed Outputs}) = 0
$$
\vskip-3pt

or even more precisely

\vskip-5pt
\vskip-5pt
$$
{\cal P}(a,b,c,d,e,f,g,h,\ldots) =
{\cal P}(b,c,d,F+i,f,g,h,F+Z1+e,\ldots)
$$
\vskip-3pt

%\noindent
%At this stage expressions of type Z1 or Z3 are placeholders
%for degree 6 polynomials yet to be specified fully.
\end{defi}
\vskip-3pt
\vskip-3pt

%The main unknown in FE is a Boolean function $Z$ and in simple cases the FE can be of type $fZ=g$ where $f$ and $g$ are two
%polynomials\footnote{Such equations have numerous non-trivial solutions, cf. \cite{toyolili}.}.
where $Z1-Z4$ will be later replaced by Boolean functions $Z(),Y(),X(),W()$. % with 6 additional inputs each.

{\bf Alternative Notation.}
There is also another notation which is more like notations used in classical invariant theory.
Instead of writing

\vskip-3pt
\vskip-3pt
$$
FE=
{\cal P}(\mbox{Inputs}) +
{\cal P}(\mbox{Transformed Outputs})
$$
\vskip-3pt

we can also write:

\vskip-6pt
\vskip-6pt
$$
FE = {\cal P} + {\cal P}^{\phi}
$$
\vskip-2pt

where

\vskip-6pt
\vskip-6pt
$$
{\cal P}^{\phi} \stackrel{def}{=} {\cal P}(input\cdot \phi) = {\cal P}(\mbox{Transformed Outputs})
$$
\vskip-2pt

which is the same as above, and we can also write:

\vskip-3pt
\vskip-3pt
$$
{\cal P}^{i}\stackrel{def}{=}{\cal P}
$$
\vskip-2pt

\vskip-3pt
\vskip-3pt
$$
{\cal P}^{o}\stackrel{def}{=}{\cal P}(input\cdot \phi)
$$
\vskip-2pt

where $\phi$ is the transformation induced by 1 round of encryption.
This usage of exponents is similar as in the mathematical (Hilbertian) invariant theory.
Our exponents can be simply interpreted as transformations on polynomials,
or more precisely as operations belonging to a certain group of transformations
acting on a set of Boolean polynomials ${\cal P}$ or $A$ or other say $(azM+b) \in B_{36}$ where $B_{36}$ is the
precise ring of all Boolean polynomials in 36 variables named $a-z$ and $M-V$ as in this paper.
The notation ${\cal P}^{\phi}$ is very elegant and unhappily {\bf ambiguous} in general, because in general $\phi$ depends also on $F$ and various key bits.
Then it happens that ${\cal P}^{\phi}$ is likely to be unique nevertheless:
we are aiming at computing ${\cal P}^{\phi}$ primarily and precisely in cases where the result,
the transformed and substituted polynomial ${\cal P}^{\phi}$ is such
that the final result ${\cal P}^{\phi}$ does NOT depend on $F,K,L$.

\noindent
At this stage expressions of type Z1 later renamed as $Z()$ or $Z3$ which will be later replaced by a function $X()$
are placeholders for degree 6 polynomials yet to be specified fully.

\vskip-6pt
\vskip-6pt
\subsection{Less Compact Notations - Expanding the Fundamental Equation}
%\vskip-1pt

The main unknown in FE is a Boolean function $Z$ and in some very simple cases the FE
can be of type $fZ=g$ where $f$ and $g$ are two
polynomials\footnote{Such equations have numerous non-trivial solutions, cf. \cite{toyolili}.}.
In general however this equation is very complex and maybe has no solutions $Z$ whatsoever.
In order to study this question in detail, in the next step, $Z$ will be represented by
its Algebraic Normal Form (ANF) with 64 binary variables which are the coefficients of the ANF of $Z$,
and there will be several equations,
and four {\bf instances} $Z1-Z4$ renamed as $Z,Y,X,W$ of the same $Z$:

\begin{defi}[A Multivariate FE]
\label{defiRewriteFEZ00}
At this step %Furthermore
we will rewrite FE as follows. We will replace Z1 by:

\vskip-6pt
\vskip-6pt
$$
%Z \leftarrow Z00+Z01*L+Z02*c+Z03*Lc+\ldots +Z62*cklfh+Z63*Lcklfh
%L,j,h,f,p,d
Z \leftarrow Z00+Z01*L+Z02*j+Z03*Lj+\ldots +Z62*jhfpd+Z63*Ljhfpd
$$
\vskip-1pt

%\medskip
Likewise we will also replace $Z2$:
\vskip-6pt
\vskip-6pt
$$
Y \leftarrow Z00+Z01*k+Z02*l+Z03*kl+\ldots +Z62*loent+Z63*kloent
$$
\vskip-1pt
\noindent
and likewise for $X=Z3$ and $W=Z4$ and the coefficients $Z00\ldots Z63$
will be the same inside $Z1-Z4$, however the subsets of 6 variables
chosen out of 36 will be different in $Z1-Z4$.
Moreover, some coefficients of ${\cal P}$ may also be variable.
\end{defi}
\vskip-3pt

%{\bf Note.} Our compact notations omit the stars for products of small variables.

%%\vskip-6pt
%%\vskip-6pt
%\subsection{On Choice of ${\cal P}$ in our FE}
%\vskip-1pt
%
%Initially, we can select ${\cal P}$ as an arbitrary fixed polynomial, with degree say between 2 and 26.
%For example one we have chosen ourselves(!) or one previously seen to work for that cipher,
%or one based on well-known symmetric polynomials with some unknowns, cf. Section \ref{FirstGlimpseOnPSymmetries}.
%Then if we cannot find a solution,
%we will enlarge the space of solutions but making more
%or all coefficients of ${\cal P}$ variable.
\noindent
In all cases, all we need to do is to solve the equation above for $Z$,
plus a variable amount of extra variables e.g. $Z63$.
This formal algebraic approach, if it has a solution,
still called $Z$
for simplicity, %in order to simplify,
or $({\cal P},Z)$
will {\bf guarantee} that our invariant ${\cal P}$ holds for 1 round.
%The key observation is that we can actually {\bf guarantee} that ${\cal P}$ is an invariant
%attack on our cipher by solving this relatively simple algebraic equation
%in order to determine the unknown Boolean function $Z$ and the coefficients of the
%polynomial ${\cal P}$. Then we win if
This is, and in this paper we are quite lucky,
IF this equation does not depend on three bits $F,K,L$.
%Potentially there is no more need for checking countless
This is the discovery process of \cite{BackdTut}
which we do not use here. We rather work with basic paper and pencil maths
and build our attack from scratch in stages. % without using a computer.

\newpage

\vskip-8pt
\vskip-8pt
\section{A New Invariant Attack of Degree 7}
\label{Constr26XByHandStage1ProductAllDeg7ver11}
\vskip-4pt

Let
\vskip-5pt
\vskip-5pt
$$
%\left\lbrace
\begin{cases}
A\stackrel{def}{=}  (i + m)~~~~\mbox{which is bits}~24,28
\cr
B\stackrel{def}{=}  (j + n)~~~~\mbox{which is bits}~23,27
\cr
C\stackrel{def}{=}  (k + o)~~~~\mbox{which is bits}~22,26
\cr
D\stackrel{def}{=}  (l + p)~~~~\mbox{which is bits}~21,25
\cr
E\stackrel{def}{=}  (y + O)~~~~\mbox{which is bits}~8,12
\cr
F\stackrel{def}{=}  (z + P)~~~~\mbox{which is bits}~7,11
\cr
G\stackrel{def}{=}  (M + Q)~~~~\mbox{which is bits}~6,10
\cr
H\stackrel{def}{=}  (N + R)~~~~\mbox{which is bits}~5,9
\cr
\end{cases}
%\right.
$$
\vskip-1pt

\vskip-6pt
\vskip-6pt
\begin{figure}
\vskip-6pt
\hskip-10pt
\hskip-10pt
\begin{center}
\hskip-10pt
\hskip-10pt
\vskip-6pt
\vskip-6pt
\vskip-6pt
\includegraphics*[width=5.0in,height=0.36in]{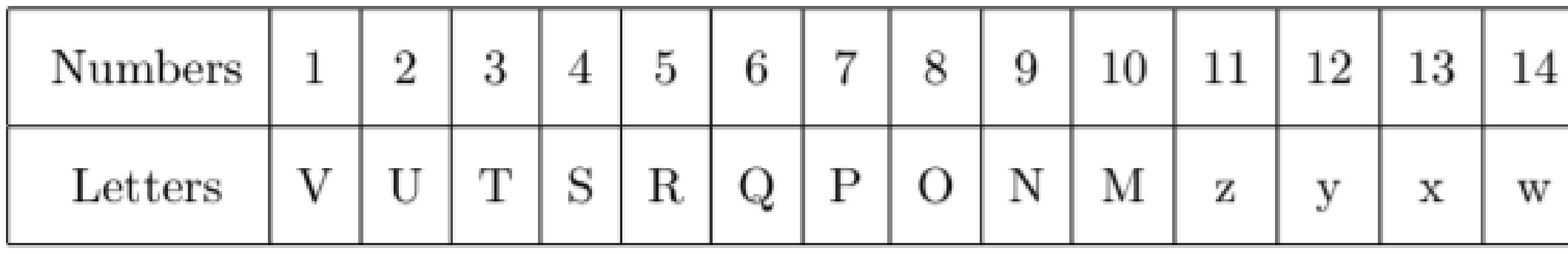}
\end{center}
\vskip-8pt
\vskip-8pt
\caption{Variable naming conventions}
\label{convert}
\end{figure}
\vskip-5pt
\vskip-5pt

\begin{theor}[A Degree 7 Invariant Attack]
\label{ProductEightThm26XGenThm7ver11}
Let
\vskip-9pt
\vskip-9pt
$$
{\cal P}=(A + B)~(C + D)~(D + F)(B + F)~(E + F)(G + F)(G + H)
%A=m+i+n+j B=o+k+p+l C=p+l+P+z D=n+j+P+z E=O+y+P+z F=P+z+Q+M G=Q+M+R+N
%AB CD  BDEG with F and FH with G where F+G counted twice
%, confirmed see degree7_P_022019.mws
%
$$
then
$\cal P$ is a non-zero polynomial of degree 7.
We also assume that
\vskip-3pt
\vskip-3pt
$$
\begin{cases}
\{D(2),D(3)\}=\{6\cdot 4,7\cdot 4\}\cr
\{D(6),D(7)\}=\{2\cdot 4,3\cdot 4\}\cr
\end{cases}
$$
and
that inputs of $Y$ are in order bits
$27,6,10,23,21,25$
and inputs of $W$
are in order bits $26,9,5,22,7,11$.
We assume that the Boolean function used inside the cipher
has after adding 1 TWO degree 3 annihilators as follows:
%ax64 587587 b+ac+bc+abc+bd+abd+bcd+abcd+e+ce+ace+bde+af+bf+abf+bcf+df+cdf+abcdf+ef+bef+cef+acef+bcef+bcdef+abcdef+1
\begin{verbatim}
(Z+1)*(f+e)(d+a)(b+c)=0
(Z+1)*(f+e+1)(d+a+1)(b+c+1)=0
\end{verbatim}
%degree7_Courtois_11022019_split_Boolean.mws
Then ${\cal P}$ is a round invariant for
any key any IV and any number of rounds.
\end{theor}

\noindent{\bf Remarks:}
We describe an attack initially designed for LZS 265 however it uses only few properties of 265,
and any key which has properties listed above will also be broken.
It may seem that Boolean functions which satisfy these properties are extremely rare.
In reality annihilators of degree 3 are near-systematic cf. Thm. 6.0.1. in \cite{toyolili},
and annihilators of degree 3 of a very special form required are also quite frequent, see \cite{BackdAnn}.
Therefore this attack CAN happen when a Boolean function chosen at random (this is how it was found in the first place).
One example of a Boolean function which works is
$b+ac+bc+abc+bd+abd+bcd+abcd+e+ce+ace+bde+af+bf+abf+bcf+df+cdf+abcdf+ef+bef+cef+acef+bcef+bcdef+abcdef+1$.
Moreover there exist further transposed variants of this attack where the inputs of $Z$ are negated
and $\cal P$ need to be modified likewise.
Moreover there exist numerous permutations of inputs of $Y$ and $W$ which will also work.
For example $b,c$ could be swapped or exchanged with the pair with $e,f$ in any order.
However even the basic attack is very complex to explain and it is hard to see if all the stages are correct.
This is why we present only the basic attack (carefully verified step by step with computer algebra software).
A full extended attack with additional cases will be published in the future.

\medskip
\noindent{\bf Proof of Thm. \ref{ProductEightThm26XGenThm7ver11}}
%\noindent\emph{Proof:}
We use the same notations as in \cite{BackdAnn}
%Thm. \ref{ProductEightThm26X} and initially
and we distinguish input and output-side variables and polynomials by $A^o$ vs. $A^i$.
We recall our assumption:
\vskip-3pt
\vskip-3pt
$$
\begin{cases}
\{D(2),D(3)\}=\{6\cdot 4,7\cdot 4\}\cr
\{D(6),D(7)\}=\{2\cdot 4,3\cdot 4\}\cr
\end{cases}
$$
\vskip-4pt
and following \cite{BackdAnn} % Section \ref{Constr26XByHandStage1FKL}
or simply following step by step a walk from output 9 to output 5 on
Fig. \ref{FigComplicationUnit6basic} above we see that:

\setcounter{equation}{0}
\vskip-9pt
\vskip-9pt
\begin{align}
H^o=y_{9}+y_{5}= x_{D(3)} + W(.) + x_{D(2)}=W(.)+A^i\\
D^o=y_{25}+y_{21}= x_{D(7)} + Y(.) + x_{D(6)}=Y(.)+E^i
\end{align}
\vskip-1pt
\vskip-1pt

%which are again seen as showing how the polynomials $D$ and $H$ on the output side
%can be rewritten as expressions using only input-side variables (where all bits $FKL$ are already eliminated).
%Following Def. \ref{defiCompactFE} all we have to do is to add the polynomial $\cal P$ to itself on the output side
%after substitutions with input-side variables.
Then we remark that some transitions are trivial
for example $H^i=G^i$ and many other:
$H \to G \to F\to E$  and $D \to C \to B \to A$.
The output polynomial is then equal to:

\vskip-7pt
\vskip-7pt
$$
(A^o + B^o)~(C^o + D^o)~(D^o + F^o)(B^o + F^o)~(E^o + F^o)(G^o + F^o)(G^o + H^o)
$$
\vskip-6pt
\vskip-6pt
\vskip-6pt
$$
(B^i + C^i)~(D^i + Y(.)+E^i)~(Y(.)+E^i + G^i)(C^i + G^i)~(F^i + G^i)(H^i + G^i)(H^i + W(.)+A^i)
$$
\vskip-3pt
at this moment we have only inputs left and we can use shorter notations:

%\vskip-6pt
%\vskip-6pt
%$$
%(B+C)~(D+Y(.)+E)~(Y(.)+E+G)(C+G)~(F+G)(H+G)(H+W(.)+A)
%%(B+I)(C+I)(D+I)(F+J)(G+J)(H+J) (YW+YA+EW+EA)
%$$
%\vskip-1pt

\setcounter{equation}{1}
\vskip-9pt
\vskip-9pt
\begin{align}
(B+C)(D+Y+E)(Y+E+G)(C+G)(F+G)(H+G)(H+W+A)
\end{align}
\vskip-1pt
\vskip-1pt

Finally we add the last expression to the input polynomial
$(A + B)(C + D)(D + F)(B + F)(E + F)(G + F)(G + H)$
and obtain that the our invariant hold is and only our sum two polynomials
is zero. In other terms our $FE$ sum which we would like to be zero is exactly equal to:
%(as found by UCL student Anthony Romm):

\vskip-3pt
\vskip-3pt
$$
(G+H)(F+G)\cdot [
$$
\vskip-9pt
\vskip-9pt
\vskip-9pt
$$
(A + B)(C + D)(D + F)(B + F)(E + F)+
(B+C)(C+G)(D+Y+E)(Y+E+G)(H+W+A)
$$
\vskip-7pt
\vskip-7pt
\vskip-7pt
$$
]
$$
\vskip-3pt

which is the same as:

%\vskip-3pt
%\vskip-3pt
%$$
%(G+H)(F+G)(1+B+D)(C+D)
%%\cdot
%[ADF+ADG+ADY+AE+AEF+AEG+AF+AGY+AY+DE+DEF+
%$$
%\vskip-7pt
%\vskip-7pt
%\vskip-7pt
%$$
%DEH+DEW+DGH+DGW+DHY+DWY+EGH+EGW+EH+EW+GHY+GWY+HY+WY]
%$$
%\vskip-1pt
%%maybe this factorization is toxic??? not unique!!!

%\newpage

%_Aidan Wrote 25022019 :
%validated in 2nd part of degree7_Courtois_11022019_split_Aidan.mws

\vskip-3pt
\vskip-3pt
$$
(F + G)(G + H)(C + D)(B + C)(D + F) *
$$
\vskip-7pt
\vskip-7pt
\vskip-7pt
$$
[(A + B)(E + F)(B + F) +
(A + H)(D + E)(B + F) + Y (G + D + 1)(B + F)(H + F + 1)(A + H) +
$$
\vskip-5pt
\vskip-1pt
\setcounter{equation}{2}
\vskip-9pt
\vskip-9pt
\begin{align}
W(H + F + 1)(G + D + 1)(D + E) + YW]
\end{align}
\vskip-1pt
\vskip-1pt

%the Aidan says that this is true (proof to be found)

This is more complex than previous results of this type.
The next step will be to absorb $Y$ and $W$.
We will need two trivial properties which come from
$(Z+1)*(f+e)(d+a)(b+c)=0$ where the order of the 6 variables matters
and exact definitions of A-H matter,
and this leads to the following 2 absorption properties:

\setcounter{equation}{3}
\vskip-9pt
\vskip-9pt
\begin{align}
CHF\cdot W=CHF
\mbox{~~~~and~~~~}
BDG\cdot Y=BDG
\end{align}
\vskip-1pt
\vskip-1pt

Then we have two more trivial properties
where the order of the 6 variables matters
which come from the fact that $(Z+1)*(f+e+1)(d+a+1)(b+c+1)=0$.
We get two more absorption properties:

\setcounter{equation}{4}
\vskip-9pt
\vskip-9pt
\begin{align}
(C+1)(H+1)(F+1)\cdot W=(C+1)(H+1)(F+1)
\end{align}
\vskip-1pt
\vskip-1pt

and

\setcounter{equation}{5}
\vskip-9pt
\vskip-9pt
\begin{align}
(B+1)(D+1)(G+1)\cdot Y=(B+1)(D+1)(G+1)
\end{align}
\vskip-1pt
\vskip-1pt

Now we are going to prove an intermediate result (lemma).
Let
\setcounter{equation}{6}
\vskip-9pt
\vskip-9pt
\begin{align}
\mu = (G + F)(G + H)(C + D)(B + C)(D + F)
\end{align}
\vskip-1pt
\vskip-1pt

then we have

\setcounter{equation}{7}
\vskip-9pt
\vskip-9pt
\begin{align}
Y \mu = \mu
\mbox{~~~~and~~~~}
W \mu = \mu
\end{align}
\vskip-1pt
\vskip-1pt

%confirmed see degree7_Courtois_11022019_split_AidanYW.mws

An interesting question is how do we factor multivariate polynomials.
Factorisation is not unique and not (or not yet) implemented in SAGE.
However the computation of the annihilator space is implemented. Here is a code snippet:

\vskip-2pt
\vskip-2pt
\begin{verbatim}
sage: R.<a,b,c,d,e,f> = BooleanPolynomialRing(6)
sage: F=BooleanFunction(mu)
sage: F.annihilator(1, dim=True)
\end{verbatim}
\vskip-3pt

\noindent
With this ability to find individual factors, here each time we have some a linear factor $f$, and we will select one such factor at random if there is more than one,
we can then divide by $f+1$. We repeat this process at random (there are many possible factorizations typically).
In this way we obtain the following two remarkable factorisations.
Here exactly is where {\bf the lack of unique factorization} plays an important role.
We get both that:

%Aidan email from 27022019
%When reviewing the absorption assumptions:
%
%(1) (F + G)*(G + H)*(C + D)*(B + C)*(D + F) * Y == (F + G)*(G + H)*(C + D)*(B + C)*(D + F)
%
%(2) (F + G)*(G + H)*(C + D)*(B + C)*(D + F) * W == (F + G)*(G + H)*(C + D)*(B + C)*(D + F)
%
%
%I have stumbled upon a result that I can't really explain because it seems like magic.
%
%
%Re-factoring the magic polynomial (right side for both) in two different ways yields:
%
%
%(3)

\setcounter{equation}{8}
\label{LackUniqueFactUsed26511}
\vskip-4pt
\vskip-4pt
\begin{align}
\mu =
%(BDG + H(BD + BG + B + DG + D + G + 1)) * (C + H + 1) * (C + F + 1) * (F + H + 1)
[ H(B+1)(D+1)(G+1) + (H+1)BDG ] (C + H + 1)(C + F + 1)(F + H + 1)
\end{align}
\vskip-1pt
\vskip-1pt
%correct!
%(BDG+H*sth)sth
%

\setcounter{equation}{9}
\vskip-4pt
\vskip-4pt
\begin{align}
\mu =
%=  (CHF+G(CF + CH + C + FH + F + H + 1)) * (B + D + 1) * (D + G + 1) * (B + G + 1)
[  G(C+1)(F+1)(H+1) + (G+1)CHF ) ] (B + D + 1)(D + G + 1)(B + G + 1)
\end{align}
\vskip-1pt
\vskip-1pt
%correct!
%(CHF+G*sth)sth

%============================================================================
%key insight: lack of unique factorization helps the attacker
%algebraic complexity reduction without unique factorization: increased power!
%degree 7 product is harder, fast phase transition, degree 12 could be incredibly easy!
%============================================================================

these two facts imply that both $Y$ and $W$ can be absorbed by $\mu$.
More precisely in (9) we see that $Y$ is absorbed by the first factor using (6) and (4)
therefore $Y \mu = \mu$.
Similarly in (10) $W$ is absorbed by the first factor using (5) and (4)
therefore $W \mu = \mu$.
We have eventually proven the result claimed earlier:
\setcounter{equation}{7}
\vskip-9pt
\vskip-9pt
\begin{align}
Y \mu = \mu
\mbox{~~~~and~~~~}
W \mu = \mu
\end{align}
\vskip-1pt
\vskip-1pt

\noindent
In the previous formula we are now allowed to replace $Y$ and $W$ by 1 (and $YW$ also by $1$).
This is because the whole expression has $\mu$ as a factor, which term absorbs both $Y$ and $W$.
%For example for any polynomial $\cal R$ inside the expression:
%$$
%\mu \cdot R)
%$$
In other words, the previous (3) becomes (11) where we replaced $Y$ and $W$  by 1
(due to absorbtion by $\mu$ which is the first factor):

\vskip-3pt
\vskip-3pt
$$
%(F + G)(G + H)(C + D)(B + C)(D + F) *
\mu \cdot
$$
\vskip-7pt
\vskip-7pt
\vskip-7pt
$$
[(A + B)(E + F)(B + F) +
(A + H)(D + E)(B + F) + (G + D + 1)(B + F)(H + F + 1)(A + H) +
$$
\vskip-5pt
\vskip-1pt
\setcounter{equation}{10}
\vskip-9pt
\vskip-9pt
\begin{align}
(H + F + 1)(G + D + 1)(D + E) + 1]
\end{align}
\vskip-1pt
\vskip-1pt

%degree7_Courtois_11022019_split_Aidan.mws

\noindent
At the end it becomes a purely syntaxic process: as $W$ and $Y$ are no longer present.
It is now sufficient to verify with formal algebra software that this polynomial is zero.
Moreover at this stage definitions of A-H do no longer matter and there result holds more generally.
%We can show this for any 8 Boolean variables.
In SAGE maths we just need to type: %declare the 8 variables as follows:

%P*(Q+1)=0
%because
%PQ=P
%confirmed see degree7_Courtois_11022019_split_Aidan.mws
%Q:=1+RemovePowers(expand( (A + B)*(E + F)*(B + F) +
%> (A + H)*(D + E)*(B + F) + (G + D + 1)*(B + F)*(H + F + 1)*(A + H) + (H + F + 1)*(G + D + 1)*(D + E) + 1 )) mod 2;
%P:=(G + F)*(G + H)*(C + D)*(B + C)*(D + F);

\vskip-2pt
\vskip-2pt
\begin{verbatim}
sage: B8.<A,B,C,D,E,F,G,H> = BooleanPolynomialRing()
sage: mu = (F + G)*(G + H)*(C + D)*(B + C)*(D + F)
sage: f = (A + B)*(E + F)*(B + F) + (A + H)*(D + E)*(B + F) +
(G + D + 1)*(B + F)*(H + F + 1)*(A + H) +
(H + F + 1)*(G + D + 1)*(D + E) + 1
sage: mu*f
0\end{verbatim}
\vskip-3pt

%this allows to do the reductions $A^2=A$ automatically.
This verification ends our proof.
Formal annihilation of the XOR of the input and output polynomials
means their formal equality for any input key and IV bit,
and insures that we have found an invariant attack on our block cipher.

%_Aidan Wrote earlier :
%
%Taking the fundamental equation, and stripping away to all the terms that contain Y,
%and then factoring out all the non-relevant variables, we get
%
%Y*(CFH + CF + CH + C + FH + F + H + 1)*(non-relevant variables)
%%only the parts which are multiples of Y
%
%Complete the same process for W, and we get
%
%W*(BDG + BD + BG + BG + B + DG + G + 1)*(non-relevant variables)
%%only the parts which are multiples of W
%
%
%These two polynomials are, after substitution, identical w/ reference to the respective Boolean functions W, Y, and are absorbers. There are terms with both W and Y, but after splitting they are both absorbed (not tested yet!).
%
%Showing that W, and Y are absorbed in this way allows us to use the following result
%
%
%if W*P = P, where W*P describes all terms with W in them, then we can set W = 1 in the fundamental equation.
%
%The same applies to Y.
%
%Upon setting W, Y = 1, It quickly becomes apparent that FE = 0
%_NOT what we need?
%

%which is the exact result we need:
%
%\vskip-1pt
%\vskip-1pt
%$$
%%FE=
%\left(
%\right)
%$$
%\vskip-1pt
%
%see degree7_Courtois_11022019_split.mws

\newpage

\vskip-8pt
\vskip-8pt
\section{Conclusion}
\label{Conclusion}
%\vskip-4pt
%\noindent
%%
%Extremely few attacks in symmetric cryptanalysis ever invented
%work at all when the number of rounds is very large.

Most cryptographic attacks are extremely simple, regular
%from the combinatorial point of view and are
and easily decomposed into simpler events.
This paper is about the existence of more sporadic %highly
irregular attacks on block ciphers.
For example we show %the existence of
a complex irregular linear property with a large prime period of 127 rounds
% which is a prime,
cf. Section \ref{IntroLin127}.
The next step is the study of non-linear invariants which is substantially harder.
We need to work in a white-box way with polynomial algebra.
Can we find complex invariant properties which arise ``ex-nihilo'' and are not combinations of simpler invariants? % properties?
%%% full version maybe %%%
%In order to approach this question
We study conditions under which polynomial invariant attacks are possible:
two polynomials must be equal.
%A first impression when reading this paper is that it is
This might give an impression that
%this paper is about how to
we try to ``backdoor'' a block cipher:
%how to
showing how to make it weak on purpose.
%\cite{KennyImprimitive,invglc,WhatDESay,gostdc2bHistGoodSboxes,ItalianNLBackdoorsBlock,BannierPartBack,MaliciousSHA,MaliciousKeccak}.
%\cite{KennyImprimitive,invglc,WhiteningParadox,WhatDESay,gostdc2bHistGoodSboxes,ItalianNLBackdoorsBlock,BannierPartBack,MaliciousSHA,MaliciousKeccak}.
In fact %This first impression is rather incorrect: what
we do rather ``proper'' cryptanalysis of block ciphers.
This is because absolutely every property\footnote{Such as assumptions on $P$ and $D$,
or the fact that the Boolean function will have a cubic annihilator or absorber of some peculiar form.}
we study about the cipher wiring happens with a relatively large probability.
It may happen accidentally, also when the whole cipher specification
and the non-linear components are the strongest possible and were not chosen by the attacker.
%Moreover if our polynomial invariant ${\cal P}$ of degree 7 does not work, another similar property could eventually work.
Then if our polynomial invariant ${\cal P}$ of degree 7 does not work, another one could work.
Furthermore, as the degree of ${\cal P}$ increases, the power of our attack increases,
and the success probabilities go up, for example the previous attack of degree 8 described in \cite{BackdAnn}
requires just one degree 3 special annihilator event on $Z$,
two of which are required in the new attack of degree 7 described in this paper.
%However our new attack here is less trivial and more complex. %surprising.
It remains an open problem which degree, maybe 12, is required to break some real-life versions of this cipher.
Due to the phase transition law (observed empirically) we %would
expect %some better
further favourable outcomes at higher degrees.

This paper shows that there exist complex periodic polynomial invariant attacks on block ciphers
where the polynomial is a well chosen product of 7 linear polynomials.
This sort of attacks were unthinkable in block cipher cryptanalysis even few months ago.
Our new product construction is a unique attack with a product of 7 polynomials, unlike any other attack seen before
and very surprising also when compared to the other recent product attack in \cite{BackdAnn}
based on two cycles of length 4 closely following natural transitions in a Feistel cipher with 4 branches.
Our attack seems to be of completely different nature.
We believe that the primary reason why such attacks can be made to work shifts to deeper properties and the structure
of the ring of multivariate polynomials $B_{n}$ with lack of unique factorization.
This ring offers numerous opportunities to make complex polynomials with a large number of variables (e.g. 16) disappear totally
(annihilation events), or to obtain the same polynomials in different ways (absorbtion events, non-unique factorisation events).
In our attack we have managed to make two different products of 7 factors equal (as polynomials).
%(where in one all terms are linear)
A non-trivial solution of this sort would be rather impossible with unique factorization.
We conjecture that if we want to discover further new types of non-trivial invariant attacks on block ciphers
%%% full version maybe %%%
%(and maybe also establish some impossibility results)
the attention needs to be brought primarily to study of these types of polynomial cancelation/absorption/equality/factoring events.

%\newpage

%\vskip-9pt
%\vskip-9pt

\appendix

\section{More Invariant Attacks at Degree 7}

This is not the only invariant attack of degree 7 we can construct by the methods of this paper.
For example it is possible to see that for the same LZS 265, and for the same definitions of $A,B,$ etc, %so somewhat on demand,
and for a different Boolean function

\vskip-7pt
\vskip-7pt
$$
Z=b+ac+bc+abc+bd+abd+bcd+abcd+e+ce+ace+bde+af+bf+abf+bcf+df+cdf+
$$
\vskip-6pt
\vskip-6pt
\vskip-6pt
$$
abcdf+ef+bef+cef+acef+bcef+bcdef+abcdef+1
$$
\vskip-3pt
which was also selected essentially at random with some amount of trial an error,
we have the following invariant property:

\vskip-7pt
\vskip-7pt
$$
{\cal P}=
(1+A+H)(B+H)(1+C+H)(D+H)(E+H)(1+F+H)(G+H)
$$
\vskip-3pt

This is very similar to our attack of Thm. \ref{ProductEightThm26XGenThm7ver11}
where we had:

\vskip-4pt
\vskip-4pt
$$
{\cal P}=(A + B)(C + D)(D + F)(B + F)(E + F)(G + F)(G + H)
$$
\vskip-3pt

There are indeed many similarities and some important differences.
In the first polynomial the letter $H$ appears in all terms. In the second one no letter appears everywhere.
Therefore we do {\bf not} expect the two attacks to be isomorphic:
identical modulo some permutation of the inputs of $Z$ and negating some inputs.
However we expect that they are similar at another level,
and that the success probability when the Boolean function is chosen at random
is more or less identical in both attacks due to
similar types of degree 3 annihilators for $Z$.  % but not the same.

{\bf If One Attack Fails...}
More importantly we expect that the sets of Boolean functions for which each attack works are essentially disjoint,
and different invariants will break them in different cases,
opening the possibility of breaking this cipher also with the exact Boolean function
which was specified in the 1970s and used in real-life encryption.  %government communications.

{\bf Phase Transitions: From Hard to Easy}.
We conjecture that for every LZS wiring such as 265, there exist a certain degree
$D\approx 8$ such that the proportion of Boolean functions
for which the cipher is broken by some invariant polynomial of degree $D$ is not negligible.
We expect that this proportion grows with $D$ and approaches $1$ when $D\to 36$.
Finally we expect that when maybe say when $S\geq 12$ the attacks get yet stronger, %particularly strong,
cf. the concept of (rapid) ``Phase Transition'' in Section \ref{Conclusion} and in  Section 2.4 in \cite{BackdAnn}.

\newpage

\end{document}